\documentclass[aps,prd,amsmath
,eqsecnum            % Adds section numbers to equation numbers.
,preprintnumbers
,nofootinbib         % Stops RevTeX moving footnotes to the references/bibliography
 ,showpacs          % Makes the (required) PACS numbers print out
 ,showkeys          % Makes the (required) keywords print out
,twocolumn         % for two columns per page, as in the journal.  Only do this when close to submission.
]{revtex4}
\usepackage[dvips]{graphicx}% For importing graphics: .ps, .jpg, .bmp, .pcx, & others but NOT .gif .
\usepackage{amsmath}
\usepackage{amsfonts}
\usepackage{amssymb}
\usepackage{longtable}   % For tables that are too long for 1 page.

\allowdisplaybreaks[4]     % allows page breaks in the middle of an equation group.

\newcommand{\df}{\ {\overset {\rm def} =}\ }
\newcommand{\dr}[2]{\frac {{\rm d} {#1}} {{\rm d} {#2}}}
\newcommand{\pdr}[2]{\frac {\partial {#1}} {\partial {#2}}}
\newcommand{\dril}[2]{{{\rm d} {#1}} / {{\rm d} {#2}}}

\begin{document}

\title{Geometry of the quasi-hyperbolic Szekeres models}

\author{Andrzej Krasi\'nski}
\affiliation{N. Copernicus Astronomical Centre, Polish Academy of Sciences, \\
Bartycka 18, 00 716 Warszawa, Poland} \email{akr@camk.edu.pl}

\author{Krzysztof Bolejko}

\affiliation{Sydney Institute for Astronomy, School of Physics  A28, \\
The University of Sydney, NSW 2006, Australia}
\email{bolejko@physics.usyd.edu.au}

\date {}

\begin{abstract}
Geometric properties of the quasi-hyperbolic Szekeres models are discussed and
related to the quasi-spherical Szekeres models. Typical examples of shapes of
various classes of 2-dimensional coordinate surfaces are shown in graphs; for
the hyperbolically symmetric subcase and for the general quasi-hyperbolic case.
An analysis of the mass function $M(z)$ is carried out in parallel to an
analogous analysis for the quasi-spherical models. This leads to the conclusion
that $M(z)$ determines the density of rest mass averaged over the whole space of
constant time.
\end{abstract}

\maketitle

\section{Motivation}

Continuing the research started in Refs. \cite{HeKr2008} and \cite{Kras2008},
the geometry of the quasi-hyperbolic Szekeres models is investigated. Unlike the
quasi-spherical Szekeres models that have been extensively investigated
\cite{Szek1975a} -- \cite{KrBo2012} and are rather well understood by now, the
quasi-plane and quasi-hyperbolic models are still poorly explored. This
situation has somewhat improved recently: in Ref. \cite{HeKr2008} a preliminary
investigation of the geometry of both these classes was carried out, and in Ref.
\cite{Kras2008} it was shown that the physical interpretation of the plane
symmetric models becomes clearer when a torus topology is assumed for the orbits
of their symmetry.

The present paper is an attempt to understand the geometry of the
quasi-hyperbolic model. In Sec. \ref{intszek}, the full set of the $\beta,_z
\neq 0$ Szekeres solutions is presented. In Sec. \ref{specprop} limitations for
the arbitrary functions in the quasi-hyperbolic models are discussed that result
from the spacetime signature and from the evolution equation. It is also shown
that a set where the mass function is zero is allowed to exist. In Sec.
\ref{nohor}, it is repeated after Ref. \cite{Kras2008} that the quasi-hyperbolic
Szekeres manifold is all contained within an apparent horizon, i.e., is globally
trapped. In Sec. \ref{subspgeom}, the geometry of various 2-dimensional surfaces
in the hyperbolically symmetric subcase is investigated and illustrated with
graphs. In Sec. \ref{diffshape}, it is shown what deformations to the surfaces
of constant $t$ and $\varphi$ ensue in the general quasi-hyperbolic case. In
Secs. \ref{sphmassinterpr} -- \ref{genqsphermass} various properties of the mass
function in the quasi-spherical models are discussed, in order to prepare the
ground for an analogous discussion of the quasi-hyperbolic case. This last task
is carried out in Secs. \ref{hypsym} and \ref{quasihyp}. The purpose of this was
to identify the volume in a space of constant $t$, which could be related to the
mass $M(z)$. This goal was not achieved as intended, but it was shown that
$M(z)$ determines the density of rest mass averaged over the space of constant
time. Section \ref{summary} is a summary of the results.

The aim of this paper is to advance the insight into the geometry of this class
of spacetimes. This is supposed to be the next step after the exploratory
investigation done in Ref. \cite{HeKr2008}.

\section{Introducing the Szekeres solutions}\label{intszek}

\setcounter{equation}{0}

This section is mostly copied from Ref. \cite{Kras2008}, mainly in order to
define the notation.

The metric of the Szekeres solutions is
\begin{equation}\label{2.1}
{\rm d} s^2 = {\rm d} t^2 - {\rm e}^{2 \alpha} {\rm d} z^2- {\rm e}^{2 \beta}
\left({\rm d} x^2 + {\rm d} y^2\right),
\end{equation}
where $\alpha$ and $\beta$ are functions of $(t, x, y, z)$ to be determined from
the Einstein equations with a dust source. The coordinates of (\ref{2.1}) are
comoving, so the velocity field of the dust is $u^{\mu} = {\delta^{\mu}}_0$, and
$\dot{u}^{\mu} = 0$.

There are in fact two families of Szekeres solutions, depending on whether
$\beta,_z = 0$ or $\beta,_z \neq 0$. The first family is a simultaneous
generalisation of the Friedmann and Kantowski -- Sachs \cite{KaSa1966} models.
Since so far it has found no useful application in astrophysical cosmology, we
shall not discuss it here (see Ref. \cite{PlKr2006}), and we shall deal only
with the second family.

After the Einstein equations are solved, the metric functions in (\ref{2.1})
become
\begin{eqnarray}\label{2.2}
{\rm e}^{\beta} &=& \Phi(t, z) {\rm e}^{\nu(z, x, y)}, \nonumber \\
{\rm e}^{\alpha} &=& h(z) \Phi(t, z) \beta,_z \equiv h(z) \left(\Phi,_z + \Phi
\nu,_z\right), \\
{\rm e}^{- \nu} &=& A(z)\left(x^2 + y^2\right) + 2B_1(z) x + 2B_2 (z)y + C(z),
\nonumber
\end{eqnarray}
where $\Phi(t, z)$ is a solution of the equation
\begin{equation}\label{2.3}
{\Phi,_t}^2 = - k(z) + \frac {2 \widetilde{M}(z)} {\Phi} + \frac 1 3 \Lambda
\Phi^2,
\end{equation}
while $h(z)$, $k(z)$, $\widetilde{M}(z)$, $A(z)$, $B_1(z)$, $B_2(z)$ and $C(z)$
are arbitrary functions obeying
\begin{equation}\label{2.4}
g(z) \df 4 \left(AC - {B_1}^2 - {B_2}^2\right) = 1/h^2(z) + k(z).
\end{equation}
The mass density $\rho$ is
\begin{equation}\label{2.5}
\kappa \rho c^2 = \frac {\left(2 \widetilde{M} {\rm e}^{3\nu}\right),_z} {{\rm
e}^{2\beta} \left({\rm e}^{\beta}\right),_z}; \qquad \kappa = 8 \pi G / c^4.
\end{equation}

This family of solutions has in general no symmetry, and acquires a
3-dimensional symmetry group with 2-dimensional orbits when $A$, $B_1$, $B_2$
and $C$ are constant (then $\nu,_z = 0$). The sign of $g(z)$ determines the
geometry of the surfaces of constant $t$ and $z$ and the symmetry of the $\nu,_z
= 0$ subcase. The geometry is spherical, plane or hyperbolic when $g > 0$, $g =
0$ or $g < 0$, respectively. With $A$, $B_1$, $B_2$ and $C$ being functions of
$z$, the surfaces $z =$ const within a single space $t =$ const may have
different geometries, i.e., they can be spheres in one part of the space and
surfaces of constant negative curvature elsewhere, the curvature being zero at
the boundary -- see a simple example of this situation in Ref.
\cite{HeKr2008}.\footnote{In most of the literature, these models have been
considered separately, but this was only for purposes of systematic research.}
The sign of $k(z)$ determines the type of evolution when $\Lambda = 0$: with $k
> 0$ the model expands away from an initial singularity and then recollapses to
a final singularity; with $k < 0$ the model is ever-expanding or
ever-collapsing, depending on the initial conditions; $k = 0$ is the
intermediate case with expansion velocity tending to zero asymptotically.

The Szekeres models are subdivided according to the sign of $g(z)$ into
quasi-spherical (with $g > 0$), quasi-plane ($g = 0$) and quasi-hyperbolic ($g <
0$). The geometry of the last two classes has, until recently, not been
investigated and is not really understood; work on their interpretation was only
begun by Hellaby and Krasi\'nski \cite{HeKr2008}, and somewhat advanced for the
quasi-plane models by the present author \cite{Kras2008}. The sign of $g(z)$
imposes limitations on the sign of $k(z)$. For the signature to be the physical
$(+ - - -)$, the function $h^2$ must be non-negative (possibly zero at isolated
points, but not in open subsets), which, via (\ref{2.4}), means that $g(z) -
k(z) \geq 0$ everywhere. Thus, with $g > 0$ all three possibilities for $k$ are
allowed; with $g = 0$ only the two $k \leq 0$ evolutions are admissible ($k = 0$
only at isolated values of $z$), and with $g < 0$, only the $k < 0$ evolution is
allowed.

The quasi-spherical models may be imagined as such generalisations of the
Lema\^i{\i}tre -- Tolman (L--T) model in which the spheres of constant mass are
non-concentric. The functions $A(z)$, $B_1(z)$ and $B_2(z)$ determine how the
center of a sphere changes its position in a space $t =$ const when the radius
of the sphere is increased \cite{HeKr2002}.

Often, it is practical to reparametrise the arbitrary functions in the Szekeres
metric as follows \cite{Hell1996}. Even if $A = 0$ initially, a transformation
of the $(x, y)$ coordinates can restore $A \neq 0$, so we may assume $A \neq 0$
with no loss of generality \cite{PlKr2006}. Then let $g \neq 0$. Writing
\begin{eqnarray}\label{2.6}
&& \left(A, B_1, B_2\right) = \frac {\sqrt{|g|}} {2S} (1, - P, - Q), \quad
\varepsilon \df g / |g|,\ \ \ \ \ \  \\
&& k = - |g| \times 2 E, \quad \widetilde{M} = |g|^{3/2} M, \quad \Phi =
\sqrt{|g|} R, \nonumber
\end{eqnarray}
we can represent the metric (\ref{2.1}) as
\begin{eqnarray}
&& \hspace{-4mm} \frac {{\rm e}^{- \nu}} {\sqrt{|g|}} \df {\cal{E}} \df \frac S
2 \left[\left(\frac {x - P} S\right)^2 + \left(\frac {y - Q} S\right)^2 +
\varepsilon\right],\ \ \ \ \ \ \ \  \label{2.7}\\
&& \hspace{-4mm} {\rm d} s^2 = {\rm d} t^2 - \frac {\left(R,_z - R {\cal
E},_z/{\cal E}\right)^2} {\varepsilon + 2E(z)} {\rm d} z^2 - \frac {R^2} {{\cal
E}^2} \left({\rm d} x^2 + {\rm d} y^2\right). \nonumber \label{2.8} \\
\end{eqnarray}
When $g = 0$, the transition from (\ref{2.1}) to (\ref{2.7}) -- (\ref{2.8}) is
$A = 1/(2S)$, $B_1 = - P/(2S)$, $B_2 = - Q/(2S)$, $k = - 2E$, $\widetilde{M} =
M$ and $\Phi = R$. Then (\ref{2.7}) -- (\ref{2.8}) applies with $\varepsilon =
0$, and the resulting model is quasi-plane.

Equation (\ref{2.3}), in the variables of (\ref{2.8}), becomes
\begin{equation}\label{2.9}
{R,_t}^2 = 2E(z) + \frac {2 M(z)} R + \frac 1 3 \Lambda R^2.
\end{equation}
{}From now on, we will use this representation. The formula for density in these
variables is
\begin{equation}\label{2.10}
\kappa \rho c^2 = \frac {2 \left(M,_z - 3 M {\cal E},_z / {\cal E}\right)} {R^2
\left(R,_z - R {\cal E},_z / {\cal E}\right)}.
\end{equation}
For $\rho > 0$, $\left(M,_z - 3M {\cal E},_z/{\cal E}\right)$ and $\left(R,_z -
R {\cal E},_z/{\cal E}\right)$ must have the same sign. Note that the sign of
both these expressions may be flipped by the transformation $z \to -z$, so we
may assume that
\begin{equation}\label{2.11}
R,_z - R {\cal E},_z/{\cal E} > 0
\end{equation}
at least somewhere. In this preliminary investigation we assume that we are in
that part of the manifold, where (\ref{2.11}) holds.

In (\ref{2.7}) -- (\ref{2.8}) the arbitrary functions are
independent.\footnote{Equation (\ref{2.4}) defines $C =
\sqrt{|g|}\left[\left(P^2 + Q^2\right)/S + \varepsilon S\right]/ 2$.} However,
(\ref{2.7}) -- (\ref{2.8}) creates the illusion that the values $\varepsilon =
+1, 0, -1$ characterise the whole spacetime, while in truth all three cases can
occur in the same spacetime.

Within each single $\{t =$ const, $z =$ const$\}$ surface, in the case
$\varepsilon = +1$, the $(x, y)$ coordinates of (\ref{2.1}) can be transformed
to the spherical $(\vartheta, \varphi)$ coordinates by
\begin{equation}\label{2.12}
(x - P, y - Q) / S = \cot (\vartheta / 2) (\cos \varphi, \sin \varphi).
\end{equation}
This transformation is called a {\it stereographic projection}. For its
geometric interpretation and for the corresponding formulae in the $\varepsilon
\leq 0$ cases see Refs. \cite{HeKr2008} and \cite{PlKr2006}.

The shear tensor for the Szekeres models is \cite{PlKr2006}
\begin{eqnarray}\label{2.13}
&& {\sigma^{\alpha}}_{\beta} = \frac 1 3 \Sigma \  {\rm diag\ } (0, 2, -1, -1),
\qquad {\rm where} \nonumber \\
&& \Sigma = \frac {R,_{tz} - R,_t R,_z / R} {R,_z - R {\cal E},_z / {\cal E}}.
\end{eqnarray}
Since rotation and acceleration are zero, the limit ${\sigma^{\alpha}}_{\beta}
\to 0$ must be the Friedmann model \cite{Elli1971,PlKr2006}. In this limit we
have
\begin{equation}\label{2.14}
R(t, z) = r(z) S(t),
\end{equation}
and then (\ref{2.9}) implies that
\begin{equation}\label{2.15}
E / r^2, M / r^3\ {\rm and}\ t_B(z)\ {\rm are\ all\ constant}.
\end{equation}
However, with $P(z)$, $Q(z)$ and $S(z)$ still being arbitrary, the resulting
coordinate representation of the Friedmann model is very untypical. The more
usual coordinates result when
\begin{equation}\label{2.16}
P,_z = Q,_z = S,_z = 0,
\end{equation}
and $r(z)$ is chosen as the $z'$ coordinate. (We stress that this is achieved
simply by coordinate transformation, but writing it out explicitly is an
impossible task.) However, (\ref{2.14}) and (\ref{2.16}) substituted in
(\ref{2.8}) give the standard representation of the Friedmann model only when
$\varepsilon = +1$. With $\varepsilon = 0$ and $\varepsilon = -1$, further
transformations are needed to obtain the familiar form
\cite{HeKr2008,RoNo1968,Kras1989}.

The above is a minimal body of information about the Szekeres models needed to
follow the remaining part of this paper. More extended presentations of physical
and geometrical properties of these models can be found in Refs.
\cite{PlKr2006,HeKr2002,HeKr2008,Kras2008,Kras1997,BKHC2010}.

\section{Specific properties of the quasi-hyperbolic model}\label{specprop}
\setcounter{equation}{0}

{}From now on we consider only the case $\Lambda = 0$, $\varepsilon = -1$ and
only expanding models. The corresponding conclusions for collapsing models
follow immediately.

It was stated in Ref. \cite{HeKr2008} that the surfaces ${\cal H}_2$ of constant
$t$ and $z$ in (\ref{2.8}) in the quasi-hyperbolic case $\varepsilon = -1$
consist of two disjoint sheets. This was a conclusion from the fact that with
$\varepsilon = -1$ the equation ${\cal E} = 0$ has a solution for $(x, y)$ at
every value of $z$, and every curve that goes into the set ${\cal E} = 0$ has
infinite length. However, it will be shown in Sec. \ref{subspgeom} that the two
sheets are in fact two coverings of the same surface, also in the general
nonsymmetric case. Their spurious isolation is a property of the stereographic
coordinates used in (\ref{2.8}).

Note that with $\varepsilon = -1$ (\ref{2.8}) shows that for the signature to be
the physical $(+ - - -)$
\begin{equation}\label{3.1}
E(z) \geq 1/2
\end{equation}
is necessary, with $E = 1/2$ being possible at isolated values of $z$, but not
on open subsets. We shall also assume
\begin{equation}\label{3.2}
M(z) \geq 0
\end{equation}
for all $z$, since with $M < 0$ (\ref{2.9}) would imply $R,_{tt} > 0$, i.e.,
decelerated collapse or accelerated expansion, which means gravitational
repulsion.

In consequence of (\ref{3.1}), only one class of solutions of (\ref{2.9}) is
possible in the quasi-hyperbolic case:
\begin{eqnarray}\label{3.3}
R &=& \frac M {2E}\ (\cosh \eta - 1), \nonumber \\
t - t_B &=& \frac M {(2E)^{3/2}}\ (\sinh \eta - \eta).
\end{eqnarray}
The second of the above determines $\eta$ as a function of $t$, with $z$ being
an arbitrary parameter, and then the first equation determines $R(t, z)$.

Equations (\ref{3.1}), (\ref{3.2}) and (\ref{2.9}) with $\Lambda = 0$ imply that
${R,_t}^2 > 0$ at all $z$, i.e., there can be no location in the manifold at
which $R,_t = 0$. In particular, there exists no location at which $R = 0$
permanently. The function $R$ attains the value $0$ only at $t = t_B$, i.e., at
the Big Bang. We have, at all points where $M > 0$,
\begin{equation}\label{3.4}
\lim_{t \to t_B} R(t, z) = 0, \qquad \lim_{t \to t_B} R,_t(t, z) = \infty,
\end{equation}
but $R > 0$ at all values of $z$ where $t > t_B$. Thus, in the quasi-hyperbolic
model there exists no analogue of the origin of the quasi-spherical model or of
the center of symmetry of the spherically symmetric model. (This fact was
demonstrated in Ref. \cite{HeKr2008} by a different method.)

However, a location $z = z_{m0}$ at which $M(z_{m0}) = 0$ is not prohibited,
even though the parameter $\eta$ in (\ref{3.3}) becomes undetermined when $M = 0
\neq E$. Writing the solution of (\ref{3.3}) as
\begin{eqnarray}\label{3.5}
&& t - t_B = \frac M {(2E)^{3/2}}\ \left[\sqrt{4E^2R^2/M^2 + 4ER/M}\right. \\
&& - \left. \ln\left(2ER/M + 1 + \sqrt{4E^2R^2/M^2 + 4ER/M}\right)\right]
\nonumber
\end{eqnarray}
(where the log-term is the function inverse to $\cosh$) we see that the limit of
this as $M \to 0$ is\footnote{The result (\ref{3.6}) shows that the argument
used in deriving the regularity conditions at the center for the Lema\^{\i}tre
-- Tolman model in Ref. \cite{MuHe2001}, and repeated in Sec. 18.4 of Ref.
\cite{PlKr2006}, was incorrect. The value of the parameter $\eta$ need not be
determined at the center. However, the resulting regularity conditions are
correct because they can be derived in a different way. Once we know that $R =
0$, $M = 0$ and $R \propto M^{1/3}$ at the center, the behaviour of $E$ at the
center follows from eq. (\ref{2.9}).}
\begin{equation}\label{3.6}
\lim_{z \to z_{m_0}} \left(t - t_B\right) = \left.\frac R {\sqrt{2E}}\right|_{z
= z_{m_0}},
\end{equation}
and the same result follows from (\ref{2.9}) with $M = 0 = \Lambda$.

Note that (\ref{3.6}) implies $R,_t\left(t, z_{m_0}\right) =
\sqrt{2E\left(z_{m_0}\right)}$ -- an expansion rate independent of time. This
agrees with Newtonian intuition -- expansion under the influence of zero mass
should proceed with zero acceleration. At all other locations, where $M(z) > 0$,
the expansion rate is greater than $\sqrt{2E(z)}$, and tends to $\sqrt{2E(z)}$
only at $R(t, z) \to \infty$. However, $E(z)$ at $z \neq z_{m_0}$ may be smaller
than $E(z_{m_0})$, so the expansion rate in the neighbourhood of the $M = 0$ set
may in fact be \textit{smaller} than $\sqrt{2 E(z_{m_0})}$.

Conversely, at a location where $R,_t =$ constant (with $\Lambda = 0$),
(\ref{2.9}) implies that $M = 0$ (because $R,_t \neq 0$ in consequence of $E
\geq 1/2, M \geq 0$ and $R \geq 0$).

The set where $M = 0$ may or may not exist in a given quasi-hyperbolic Szekeres
spacetime. It should be noted that, if it exists, it is a 3-dimensional
hypersurface in spacetime, unlike the origin in the quasi-spherical models. The
latter is a 2-dimensional surface in spacetime and a single point in each space
of constant $t$ because in the quasi-spherical case $M = 0$ implies $R = 0$ via
the regularity conditions.

\section{No apparent horizons}\label{nohor}

\setcounter{equation}{0}

In Ref. \cite{Kras2008} it was shown that a collapsing quasi-hyperbolic Szekeres
manifold is all contained within the future apparent horizon, i.e., that it
represents the interior of a black hole. This is consistent with the fact that
the corresponding vacuum solution (the hyperbolically symmetric counterpart of
the Schwarzschild solution) has no event horizons and is globally nonstatic
\cite{HeKr2008}.

Here, we consider expanding models, and an addendum is needed to the result
reported above. Consider a surface of constant $t$ and $z$ in (\ref{2.8}), and a
family of null geodesics intersecting it orthogonally. As shown in Refs.
\cite{Kras2008} and \cite{HeKr2002}, the expansion scalar for this family is
\begin{equation}\label{4.1}
{k^{\mu}};_{\mu} = 2 \left|\frac {R,_z} R - \frac {{\cal E},_z} {\cal E}\right|
\left(\frac {R,_t} {\sqrt{2E - 1}} + e\right),
\end{equation}
where $e = +1$ for ``outgoing'' and $e = -1$ for ``ingoing''
geodesics;\footnote{Since the surfaces of constant $t$ and $z$ are infinite,
this labeling is purely conventional in this case, but the two families are
distinct.} eq. (\ref{4.1}) was adapted to $\varepsilon = -1$.

For an expanding model $R,_t > 0$, and only past-trapped surfaces can possibly
exist, for which ${k^{\mu}};_{\mu} > 0$. Apart from shell crossings the first
factor in (\ref{4.1}) is positive everywhere. Hence, (\ref{4.1}) implies
\begin{equation}\label{4.2}
\frac {R,_t} {\sqrt{2E - 1}} + e > 0.
\end{equation}
For $e = +1$, and with $R,_t > 0$ that we now consider, this is fulfilled
everywhere. For $e = -1$ we get
\begin{equation}\label{4.3}
{R,_t}^2 > 2E - 1 \geq 0
\end{equation}
(the last inequality from (\ref{3.1})). With $\Lambda = 0$, (\ref{4.3}) is also
guaranteed to hold everywhere, by (\ref{2.9}), since $M \geq 0$ and $R \geq 0$.
This means that every surface of constant $t$ and $z$ in an expanding model is
past-trapped at all of its points. But then, every point of the Szekeres
manifold lies within one such surface. This, in turn, means that every point of
the Szekeres manifold is within a past-trapped region. Therefore, the whole
quasi-hyperbolic expanding Szekeres manifold is within a past apparent horizon.

The fact of being globally trapped is a serious limitation on the possible
astrophysical applications of the quasi-hyperbolic model.

\section{Interpretation of the coordinates of (\ref{2.8})}\label{intcoord}

\setcounter{equation}{0}

In order to understand the geometry of (\ref{2.8}), we begin with the
hyperbolically symmetric subcase, $P,_z = Q,_z = S,_z = 0$. It is most
conveniently represented as
\begin{equation}\label{5.1}
{\rm d} s^2 = {\rm d} t^2 - \frac {{R,_z}^2 {\rm d} z^2} {2E - 1} - R^2
\left({\rm d} \vartheta^2 + \sinh^2 \vartheta {\rm d} \varphi^2\right).
\end{equation}
The two supposedly disjoint sheets of a constant-$(t, z)$ surface, in the
coordinates of (\ref{2.8}), are
\begin{eqnarray}\label{5.2}
&& {\rm sheet\ 1:} \quad \left(\frac {x - P} S\right)^2 + \left(\frac {y - Q}
S\right)^2 > 1, \nonumber \\
&& {\rm sheet\ 2:} \quad \left(\frac {x - P} S\right)^2 + \left(\frac {y - Q}
S\right)^2 < 1.
\end{eqnarray}
The transformation from sheet 1 to (\ref{5.1}) is
\begin{equation}\label{5.3}
(x, y) = (P, Q) + S \coth (\vartheta / 2) (\cos \varphi, \sin \varphi),
\end{equation}
while the transformation from sheet 2 is
\begin{equation}\label{5.4}
(x, y) = (P, Q) + S \tanh (\vartheta / 2) (\cos \varphi, \sin \varphi).
\end{equation}
This shows that the two sheets are in truth two coordinate coverings of the same
surface. The direct coordinate transformation  between the two sheets is the
inversion
\begin{equation}\label{5.5}
(x - P, y - Q) = \frac {S^2 (x' - P, y' - Q)} {(x' - P)^2 + (y' - Q)^2}.
\end{equation}

\begin{figure*}
 \includegraphics{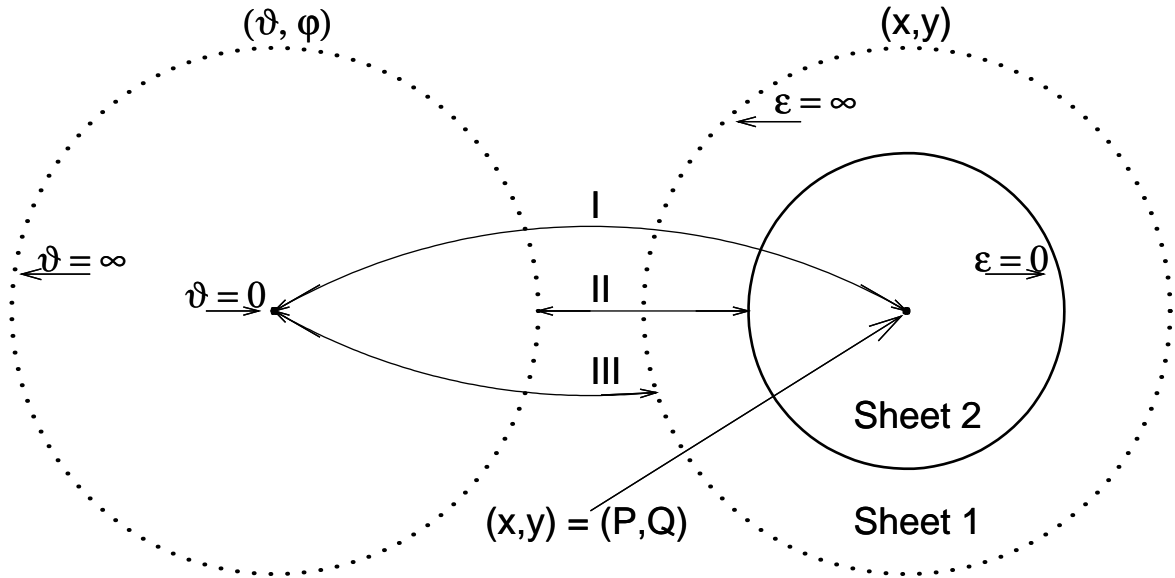}
 \caption{
 \label{maps}
 \footnotesize
Relations between the $(\vartheta, \varphi)$ and $(x, y)$ maps of a
constant-$(t, z)$ surface in (\ref{2.8}) and (\ref{5.1}). The arrow marked by I
corresponds to the transformation (\ref{5.4}) that maps the set $\vartheta = 0$
to $(x, y) = (P, Q)$. Arrow II shows that both (\ref{5.3}) and (\ref{5.4}) map
$\vartheta \to \infty$ to the circle ${\cal E} = 0$. Arrow III corresponds to
(\ref{5.3}) that maps $\vartheta = 0$ to ${\cal E} \to \infty$. }
 \end{figure*}

The circle separating the two sheets, $(x - P)^2 + (y - Q)^2 = S^2$, on which
${\cal E} = 0$, corresponds to $\vartheta \to \pm \infty$ in the coordinates of
(\ref{5.1}). The center of this circle, $(x, y) = (P, Q)$, which is in sheet 2,
is mapped by (\ref{5.4}) to $\vartheta = 0$. The infinity, $(x - P)^2 + (y -
Q)^2 \to \infty$, which is in sheet 1, is mapped by (\ref{5.3}) also to
$\vartheta = 0$. These relations are illustrated in Fig. \ref{maps}.

There is no reason to allow negative values of $\vartheta$ in (\ref{5.1})
because, as both (\ref{5.3}) and (\ref{5.4}) show, the point of coordinates $(-
\vartheta, \varphi)$ coincides with the point of coordinates $(\vartheta,
\varphi + \pi)$, so the ranges $\vartheta \in [0, + \infty)$ and $\varphi \in
[0, 2 \pi)$ cover the whole $(\vartheta, \varphi)$ surface.

Curves that go through $\vartheta = 0$ are seen from (\ref{5.1}) to have finite
length. At $\vartheta = 0$ we have $\det\left(g_{\alpha \beta}\right) = 0$, but
the curvature scalars given in Appendix \ref{appsymm} do not depend on
$\vartheta$, so $\vartheta = 0$ is only a coordinate singularity.

The geometry of the $(x, y)$ surfaces in (\ref{2.8}) is the same in the
hyperbolically symmetric case and in the full nonsymmetric case with ${\cal
E},_z  \neq 0$. In the $(x, y)$ coordinates of (\ref{2.8}), $\vartheta = 0$
corresponds to ${\cal E} = -1$ in sheet 2, which is clearly not a singularity,
and to ${\cal E} \to \infty$ in sheet 1. This seems to be a singularity in
(\ref{2.8}), but the curvature scalars are not singular there, as shown in
Appendix \ref{appnonsymm}. Also the set ${\cal E} = 0$ seems to be singular in
(\ref{2.8}), but the same formulae in Appendix \ref{appnonsymm} show that it is
nonsingular. Hence, also in the general case there is no reason to treat these
two sheets as disjoint -- they are two coordinate coverings of the same surface.

\section{Geometry of subspaces in the hyperbolically symmetric
limit}\label{subspgeom}

\setcounter{equation}{0}

\subsection{Hypersurfaces of constant $z$}

A hypersurface $z = z_1 =$ constant has the curvature tensor
\begin{eqnarray}\label{6.1}
^3R_{0 2 0 2} &=& ^3R_{0 3 0 3} / \sinh^2 \vartheta = R R,_{tt}, \nonumber \\
^3R_{2 3 2 3} &=& R^2 \sinh^2 \vartheta \left(1 - {R,_t}^2\right),
\end{eqnarray}
where $(x^0, x^2, x^3) = (t, \vartheta, \varphi)$. Consequently, it is flat when
$R,_t = \pm 1$ and curved in every other case (also when $R,_t =$ constant $\neq
\pm 1$). However, (\ref{2.9}) implies that with $R,_t = \pm 1$ we have $M = 0$
and $E = 1/2$ (recall: we consider only the case $\Lambda = 0$). Such a subset
in spacetime (if it exists) is a special case of a neck -- see the explanation
to Fig. \ref{evolwithnoedge} later in this section.

The metric of a general hypersurface of constant $z$ is
\begin{equation}\label{6.2}
{{\rm d} s_{z_1}}^2 = {\rm d}t^2 - R^2(t,z_1) \left({\rm d} \vartheta^2 +
\sinh^2 \vartheta {\rm d} \varphi^2\right).
\end{equation}
To gain insight into its geometry, we first consider its subspace given by
$\varphi = \varphi_0 =$ constant. The $z_1$ is a constant parameter within $R$
and will be omitted in the formulae below. The corresponding 2-dimensional
metric is
\begin{equation}\label{6.3}
{{\rm d} s_{z_1, \varphi_0}}^2 = {\rm d}t^2 - R^2(t) {\rm d} \vartheta^2 \equiv
\left(\dr t R\right)^2 {\rm d}R^2 - R^2 {\rm d} \vartheta^2.
\end{equation}
This can be embedded in a flat 3-dimensional Minkowskian space with the metric
\begin{equation}\label{6.4}
{{\rm d} s_M}^2 = {\rm d}T^2 - {\rm d} X^2 - {\rm d} Y^2
\end{equation}
by
\begin{eqnarray}\label{6.5}
T &=& \int \sqrt{1 + \left(\dr t R\right)^2} {\rm d} R, \nonumber \\
X &=& R \cos \vartheta, \qquad Y = R \sin \vartheta.
\end{eqnarray}
The embedding (\ref{6.5}) projects a point of coordinates $(R, \vartheta)$ and
points of coordinates $(R, \vartheta + 2\pi n)$, where $n$ is any integer, onto
the same point of the Minkowskian space (\ref{6.4}). However, these points do
not coincide in the spacetime (\ref{5.1}) -- the identification of $(R,
\vartheta)$ with $(R, \vartheta + 2\pi)$ is not allowed because the
transformation $\vartheta \to \vartheta + 2\pi$ is not an isometry in
(\ref{5.1}). Thus, the surface with the metric (\ref{6.3}) is covered by the
mapping (\ref{6.5}) an infinite number of times. This shows that a
hyperbolically symmetric geometry is a rather exotic and complicated entity. We
shall see this feature further on, while considering other surfaces.

\begin{figure}[h]
 \begin{flushleft}
 ${}$ \\[-10mm]
 \hspace{-10mm}
 \includegraphics[scale=0.6]{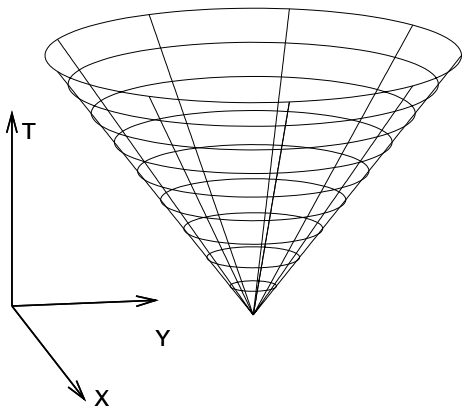}
 ${}$ \\[-10mm]
 \caption{
 \label{constzembed}
 \footnotesize
The surface of constant $z = z_1$ and constant $\varphi = \varphi_0$ in a
spacetime with the metric (\ref{5.1}). The embedding is in a Minkowskian 3-space
with the metric (\ref{6.4}). The vertex at $R = 0$ lies at the Big Bang. The
circles represent the surfaces of constant $t$ and $z$ in (\ref{5.1}). This
embedding is not a one-to-one representation, the surface in the figure is
covered with that of (\ref{6.3}) an infinite number of times -- see explanation
in the text.
  }
  \end{flushleft}
 \end{figure}

Using (\ref{2.9}) with $\Lambda = 0$ we can write
\begin{equation}\label{6.6}
\left(\dr t R\right)^2 = \frac R {2ER + 2M},
\end{equation}
and then the integral in (\ref{6.5}) can be calculated explicitly:
\begin{equation}\label{6.7}
T = \frac {FG} {2E} - \frac M {E \sqrt{2E (2E + 1)}} \ln \left(\sqrt{2E} F +
\sqrt{2E + 1} G\right) + D,
\end{equation}
where $D$ is a constant and
\begin{equation}\label{6.8}
F \df \sqrt{(2E + 1)R + 2M}, \qquad G \df \sqrt{2ER + 2M}.
\end{equation}
The constant $D$ can be chosen so that $T = 0$ at $R = 0$. Figure
\ref{constzembed} shows the graph of the surface given by the parametric
equations (\ref{6.5}) as embedded in the 3-dimensional space with the metric
(\ref{6.4}). It is not exactly a cone, the curves $T(R)$ do have nonzero
curvature
\begin{equation}\label{6.9}
\dr {^2T} {R^2} = \frac M {F G^3},
\end{equation}
but it is so small everywhere that it would not show up in a graph. Note that
the vertex angle of this conical surface is everywhere larger than $\pi / 4$,
since $\dril T R > 1$ from (\ref{6.5}).

Suppose that we made $T$ unique by choosing $D$ as indicated under (\ref{6.8}).
The vertex of the conical surface in Fig. \ref{constzembed} corresponds to the
Big Bang. If we want the image in this figure to correspond to the history of
the Universe from the Big Bang up to now, then the upper edge of the funnel
should be at $T(R_p)$, where $R_p$ corresponds to the present moment. But this
$R_p$ depends on the value of $z = z_1$. Consequently, the height of the funnel
will be different at different values of $z$.

Now we go back to (\ref{6.2}) and consider a surface of constant $\vartheta =
\vartheta_0$. Writing $C_0 = \sinh \vartheta_0$ we can write the 2-metric as
\begin{equation}\label{6.10}
{{\rm d} s_{z_1, \vartheta_0}}^2 = \left[{C_0}^2 + \left(\dr t R\right)^2\right]
{\rm d}R^2 - {\rm d} \left(C_0 R\right)^2 - \left(C_0 R\right)^2 {\rm d}
\varphi^2,
\end{equation}
and then the embedding equations are
\begin{eqnarray}\label{6.11}
T &=& \int \sqrt{{C_0}^2 + \left(\dr t R\right)^2} {\rm d} R \nonumber \\
&& \equiv C_0 \int \sqrt{1 + \frac R {2E {C_0}^2 R + 2M {C_0}^2}} {\rm d} R,
\nonumber \\
X &=& C_0 R \cos \varphi, \qquad Y = C_0 R \sin \varphi.
\end{eqnarray}
Now there is no multiple covering because $\varphi$ is a cyclic coordinate also
in spacetime, and the surface given by (\ref{6.11}) looks qualitatively similar
to that in Fig. \ref{constzembed}, except that the presence of $C_0$ introduces
some flexibility. The second line of (\ref{6.11}) shows that the explicit
expression for $T$ is (\ref{6.7}) multiplied by $C_0$, with $(M, E)$ replaced by
${C_0}^2 (M, E)$. The radius of a circle of constant $R$ is now $(C_0R)$. The
value of $C_0$ is any in $(-\infty, +\infty)$. When $C_0 \to 0$, the surface
degenerates to the straight line $X = Y = 0$. In order that $C_0 T$ in
(\ref{6.7}) allows a well-defined limit $C_0 \to 0$, the constant $D$ must have
the form
\begin{equation}\label{6.12}
D = \frac {M \ln C_0} {C_0 E \sqrt{2E \left(2E {C_0}^2 + 1\right)}} + \frac
{D_1} {C_0},
\end{equation}
where $D_1$ is another constant. Again, it may be chosen so that $C_0 T = 0$ at
$R = 0$.

{}From (\ref{6.11}) we find
\begin{equation}\label{6.13}
\lim_{C_0 \to \infty} \dr T {\left(C_0 R\right)} = 1,
\end{equation}
so in the limit $C_0 \to \infty$ the surface (\ref{6.11}) becomes exactly a cone
with the vertex angle $\pi / 4$. However, with $C_0 \to \infty$ the whole cone
recedes to infinity, as can be seen from (\ref{6.7}) and (\ref{6.11}): the
vertex of the cone, which is at $R = 0$, has the property $\lim_{C_0 \to \infty}
\left.\left(C_0T\right)\right|_{R = 0} = \infty$, even with the value of $D$
corrected as in (\ref{6.12}).

Note that the image in Fig. \ref{constzembed} will not change qualitatively when
we go over from the hyperbolically symmetric subcase (\ref{5.1}) to the general
(nonsymmetric) quasi-hyperbolic case (\ref{2.7}) -- (\ref{2.8}). Each
hypersurface of constant $z$ in it is axially symmetric, and its metric can be
transformed to the form (\ref{6.2}). Moreover, any surface of constant $z$ and
$y$ can have its metric transformed to the form (\ref{6.3}). The only change
with respect to Fig. \ref{constzembed} is that the cone-like surfaces, while
still being axially symmetric, can have their vertex angles different at
different values of $z$.

\begin{figure}[h]
 \begin{center}
 ${}$ \\[-10mm]
 \includegraphics[scale=0.5]{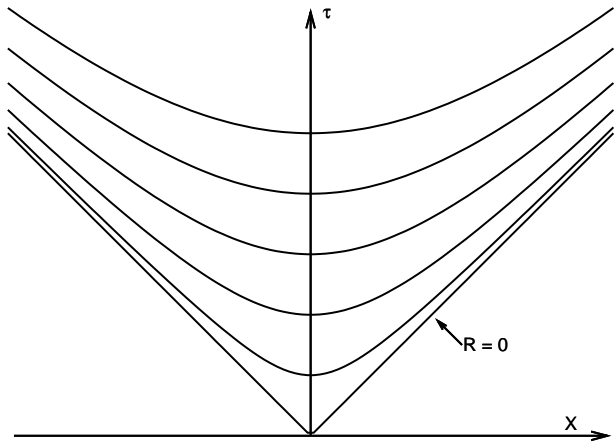}
 \caption{
 \label{hyperboloids}
 \footnotesize
An axial cross-section through the family of hyperboloids given by (\ref{6.15}).
}
 \end{center}
 \end{figure}

For completeness, we now consider the special flat hypersurface with $R,_t = 1$,
$M = 0$ and $E = 1/2$ mentioned below (\ref{6.1}). It can be all transformed to
the 3-dimensional Minkowski form. The transformation to the Minkowski
coordinates $(\tau, X, Y)$ is
\begin{eqnarray}\label{6.14}
\tau &=& R \cosh \vartheta, \qquad X = R \sinh \vartheta \cos \varphi, \nonumber
\\
Y &=& R \sinh \vartheta \sin \varphi.
\end{eqnarray}
The surfaces of constant $R$ are given by the equation
\begin{equation}\label{6.15}
\tau^2 - X^2 - Y^2 = R^2.
\end{equation}
These are two-sheeted hyperboloids when $R > 0$ and a cone when $R = 0$. They
intersect the $\tau$ axis horizontally, and all tend asymptotically to the cone
$R = 0$ as $X^2 + Y^2 \to \infty$ (see Fig. \ref{hyperboloids}).

\begin{figure}[h]
 \begin{center}
 ${}$ \\[-5mm]
 \includegraphics[scale=0.6]{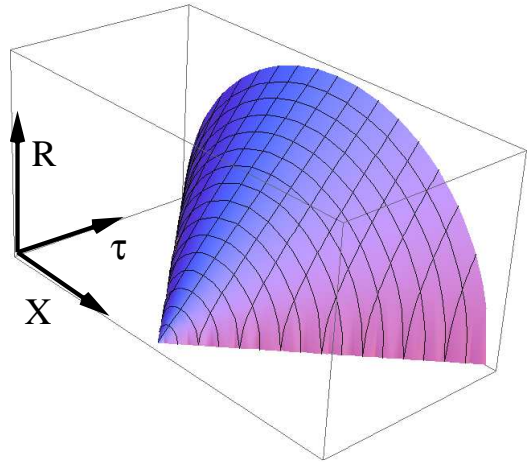}
 ${}$ \\[-10mm]
 \caption{
 \label{hypsurface}
 \footnotesize
The subspace $\{z = {\rm constant}, \varphi = 0\}$ of the spacetime (\ref{5.1})
with $R = t$. The curves of constant $X$ are the hyperbolae $\tau^2 - R^2 = X^2$
(the one with $X = 0$ is the straight line $\tau = R$). The lines of constant
$\tau$ are the circles $X^2 + R^2 = \tau^2$. }
 \end{center}
 \end{figure}

The surface $\varphi = 0$ of (\ref{6.14}) is depicted in Fig. \ref{hypsurface}.
Note, however, that Figs. \ref{hyperboloids} and \ref{hypsurface} are graphs of
a Lorentzian space mapped into a Euclidean space, so geometrical relations of
(\ref{6.2}) are not faithfully represented.

\subsection{The $R(t, z)$ curves}

\begin{figure}[h]
 \includegraphics[scale=0.3]{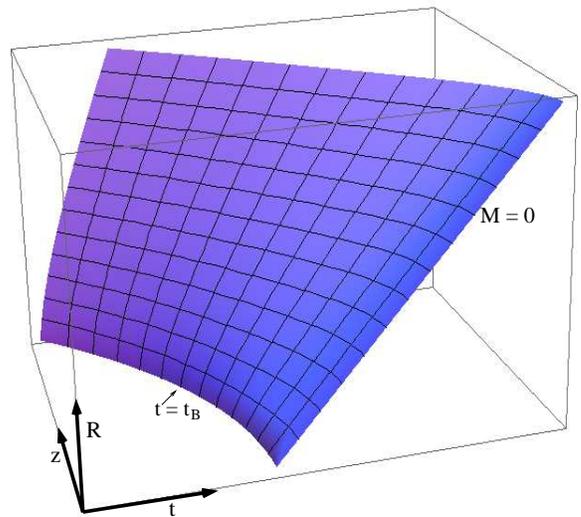}
 \caption{
 \label{evolutions}
 \footnotesize
An exemplary collection of the $R(t,z)$ curves for various fixed values of $z$.
The bang time curve $t = t_B(z)$ must be a decreasing function of $z$ to avoid
shell crossings \cite{HeKr2008}. The rightmost line is straight, corresponding
to $M = 0$. The other $t(R)$ functions in this figure are given by (\ref{3.5})
with $t_B(z) = 2 - 0.5 z^2$, $M = z^3$ and $E = 0.5 + z^{3/2}$. The values of
$z$ change by equal increments from 0 at the rightmost curve to 1 at the
leftmost curve. The curves have $M$ increasing with $z$ (so
$\left|R,_{tt}\right|$ is increasing as a function of $z$) and $E$ increasing
with $z$ (so $R,_t$ is increasing). Note that all curves except the $M = 0$ one
hit the $t = t_B$ set with $R,_t \to \infty$. The horizontal curves are those
of constant $R$; the values of $R$ on them change by equal increments from 0 on
the lowest curve to 0.8 on the highest curve.
  }
 \end{figure}

Where $M > 0$, we have $R,_{tt} < 0$ from (\ref{2.9}) with $\Lambda = 0$.
Consequently, $R$ as a function of $t$ must be concave. The slopes of the curves
$R(t, z)$ at various $z$ depend on $E(z)$, and their initial points at $t = t_B$
are determined by $t_B(z)$, so both can vary arbitrarily when we proceed from
one value of $z$ to another. Fig. \ref{evolutions} shows a 3-d graph of an
example of a family of $R(t, z)$ curves corresponding to different values of
$z$.

\subsection{Hypersurfaces of constant $t$}

Formulae for the curvature of the spaces of constant $t$ in (\ref{2.7}) --
(\ref{2.8}) are (from Ref. \cite{HeKr2008}, in notation adapted to that used
here):
\begin{eqnarray}\label{6.16}
^3R_{1212} &=& ^3R_{1313} \nonumber \\
&=& - \frac {R \left(R,_z - R {\cal E},_z / {\cal E}\right) \left(E,_z - 2E
{\cal E},_z / {\cal E}\right)} {(2E - 1) {\cal E}^2}, \nonumber \\
^3R_{2323} &=& - \frac {2E R^2} {{\cal E}^4},
\end{eqnarray}
where the coordinates are labeled as $(x^1, x^2, x^3) = (z, x, y)$. Equations
(\ref{6.16}) show that a space of constant $t$ becomes flat when $E = 0$, but
then it has the Lorentzian signature $(+ - -)$. Consequently, with the Euclidean
signature, these spaces can never be flat.

\begin{figure}[h]
 ${}$ \\[-10mm]
 \hspace*{-10mm}
 \includegraphics[scale=0.6]{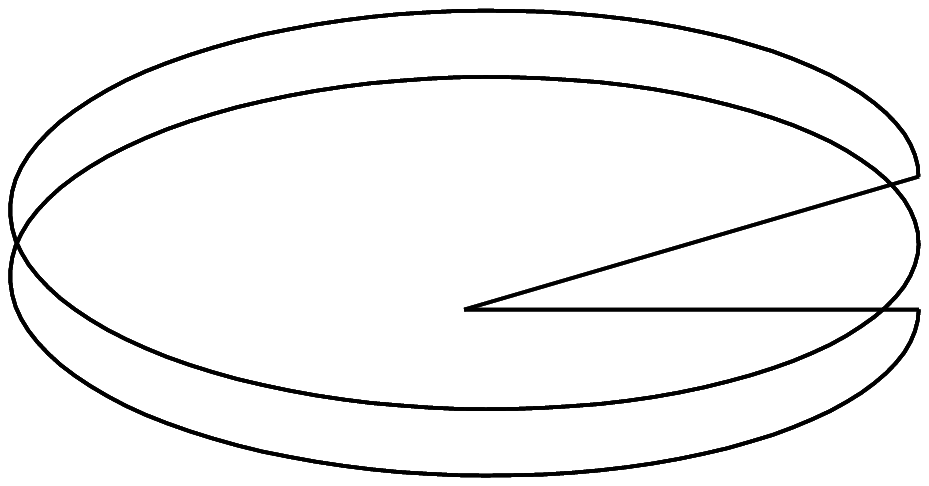}
 ${}$ \\[-20mm]
 \hspace*{-27mm}
  \includegraphics[scale=0.8]{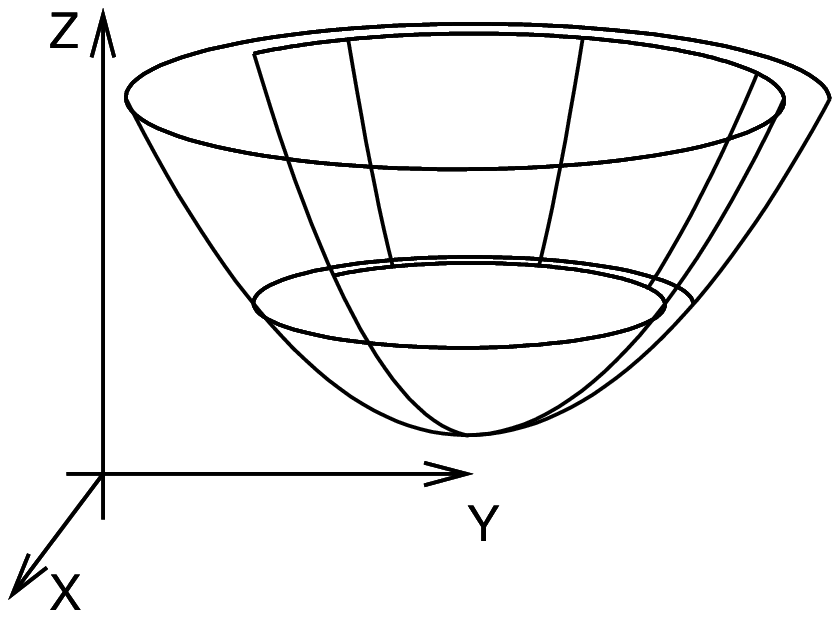}
 ${}$ \\[-10mm]
 \caption{
 \label{multisheet}
 \footnotesize
{\bf Upper graph:} A surface $H_2$ of constant $t$ and $\varphi$ in the metric
(\ref{5.1}) in the case when $E \equiv 1$. It is locally isometric to the
Euclidean plane, but points having the coordinates $(R, \vartheta + 2 \pi n)$ do
not coincide with the point of coordinates $(R, \vartheta)$, so the projection
of $H_2$ covers the Euclidean plane multiply. The multiple covering is depicted
schematically. {\bf Lower graph:} When $E$ is not constant, the embedding of a
surface of constant $t$ and $\varphi$ in the Euclidean space is locally
isometric to a curved surface of revolution, with a similar multiple covering,
also shown schematically. The surface in the figure is the paraboloid $Z = R^2$
that results when $E(z(R)) = (2R^2 + 1) / (4R^2 + 1)$, where $z(R)$ is the
inverse function to $R(t_0, z)$.  }
 \end{figure}

Now let us consider the surfaces $H_2$ of constant $t = t_0$ and $\varphi =
\varphi_0$ in (\ref{5.1}). For the beginning we will assume that $R,_z > 0$ for
all values of $z$ in the region under investigation. Then we can write the
metric of $H_2$ as follows:
\begin{equation}\label{6.17}
{\rm d} {s_2}^2 = \frac {[{\rm d} R(t_0, z)]^2} {2E - 1} + R^2 {\rm d}
\vartheta^2.
\end{equation}
When $E \equiv 1$, this is the metric of the Euclidean plane in polar
coordinates $(R, \vartheta)$. With some other constant values of $E$, this will
be the metric of a cone (see below). With other functional forms of $E$, it is
the metric of a rotationally symmetric curved surface on which $\vartheta$ is
the polar angular coordinate. We encounter here the same phenomenon that was
described in connection with Fig. \ref{constzembed}: in each case, a point of
coordinates $(R, \vartheta)$ and points of coordinates $(R, \vartheta + 2\pi
n)$, where $n$ is any integer, are projected onto the same point of the plane,
cone or curved surface, respectively. However, as before, these points do not
coincide in the spacetime (\ref{5.1}). Examples of embeddings of $H_2$ in the
Euclidean $E^3$ are illustrated in Fig. \ref{multisheet}.

Note that with $E =$ constant $\neq 1$ other interesting geometries come up. If
we interpret $\vartheta$ as a polar coordinate, then the ratio of a
circumference of a circle $R =$ constant to its radius is $2\pi \sqrt{2E - 1}$,
which means that the surfaces (\ref{6.17}) are ordinary cones when $1/2 < E <
1$, and cone-like surfaces that cannot be embedded in a Euclidean space when $E
> 1$. With $E$ being a function of $z$, the cones and/or cone-like surfaces
are tangent to the $(z, \vartheta)$ surfaces at the appropriate values of $z$.
Thus, with $E > 1$, the $(z, \vartheta)$ surfaces cannot be embedded in a
Euclidean space. Whether this surface looks like the smooth surface of
revolution in the lower panel of Fig. \ref{multisheet}, or like a cone, depends
on the behaviour of $E(z)$ in the neighbourhood of the axis $R = 0$. But
attention: if the value of $t_0$ under consideration is such that $t_0
> t_B(z)$ for all $z$, then the set $R = 0$ is not contained in the space $t =
t_0$. We will come back to this below. We can discuss the embedding when we
write the metric (\ref{6.17}) as
\begin{equation}\label{6.18}
{\rm d} {s_2}^2 = \frac {2(1 - E)} {2E - 1} {\rm d} R^2 + {\rm d} R^2 + R^2 {\rm
d} \vartheta^2.
\end{equation}
Now it is seen that with $1/2 < E < 1$ we can embed this surface in the
Euclidean space with the metric ${\rm d} {s_3}^2 = {\rm d} X^2 + {\rm d} Y^2 +
{\rm d} Z^2$ by
\begin{equation}\label{6.19}
X = R \cos \vartheta, \quad Y = R \sin \vartheta, \quad Z = \pm \int\sqrt{\frac
{2(1 - E)} {2E - 1}} {\rm d} R,
\end{equation}
while with $E > 1$ we can embed it in the Minkowskian space with the metric
${\rm d} {s_3}^2 = - {\rm d} T^2 + {\rm d} X^2 + {\rm d} Y^2$ by
\begin{equation}\label{6.20}
X = R \cos \vartheta, \quad Y = R \sin \vartheta, \quad T = \pm \int\sqrt{\frac
{2(E - 1)} {2E - 1}} {\rm d} R,
\end{equation}

For later reference let us note, from (\ref{6.19}), that $\dril Z R \to 0$ when
$E \to 1$ and $|\dril Z R| \to \infty$ when $E \to 1/2$. This observation will
be useful in drawing graphs and interpreting them.

The surfaces on which $t$ and $\vartheta$ are constant look similar to the
surfaces described above, with two differences:

1. The coordinate $\varphi$ changes from $0$ to $2 \pi$ also in the spacetime
(\ref{5.1}), so there is no multiple covering of the surfaces in the Euclidean
space.

2. The circumference to radius ratio is this time $2 \pi \sinh \vartheta
\sqrt{2E - 1}$, so the transition from cones to cone-like surfaces occurs at
$\sinh \vartheta = 1 / \sqrt{2E - 1}$.

\begin{figure}[h]
 \begin{center}
% ${}$ \\[-20mm]
% \hspace{-90mm}
  \includegraphics[scale=0.45]{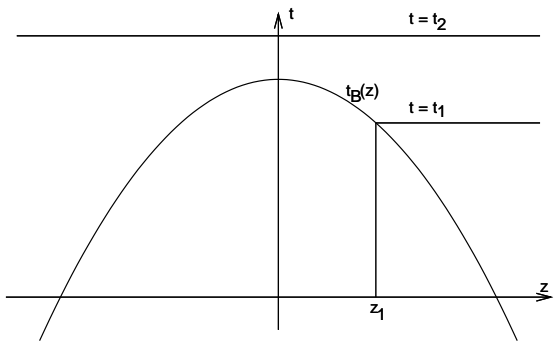}
% ${}$ \\[-30mm]
 \caption{
 \label{holenohole}
 \footnotesize
The hypersurface $t = t_1$ has a nonempty intersection with the Big Bang set $t
= t_B(z)$. The function $R(t_1, z)$ attains the value 0 at $z = z_1$, and the
surface from Fig. \ref{multisheet} extends down to the axis $R = 0$. At $t =
t_2$ we have $t > t_B$ at all values of $z$, so $R(t_2, z)$ is nowhere zero and
has a smallest lower bound $R_0 > 0$. This means that the corresponding surface
from Fig. \ref{multisheet} will have a hole of radius $R_0$ around the axis.
Since $R,_t > 0$, the radius of the hole increases with time, as shown in Fig.
\ref{multisheetexp}.}
 \end{center}
 \end{figure}

\begin{figure}[h]
 \begin{center}
 ${}$ \\[-20mm]
 \hspace*{-35mm}
  \includegraphics[scale=0.9]{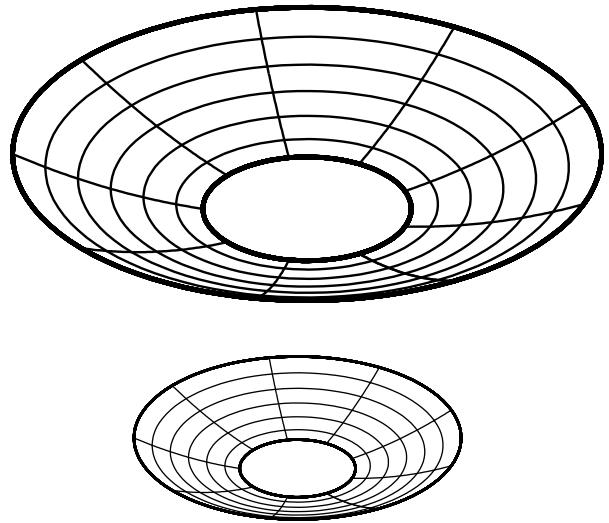}
 ${}$ \\[-40mm]
  \includegraphics[scale=0.5]{multisurfexp.ps}
 ${}$ \\[-20mm]
 \caption{
 \label{multisheetexp}
 \footnotesize
The surface of constant $t$ and $\varphi$ of (\ref{5.1}), from the bottom graph
in Fig. \ref{multisheet}, depicted at two instants $t_2 > t_B$ (bottom graph)
and $t_3 > t_2$ (top graph). The multiple covering of the paraboloid is no
longer taken into account. The hole around the axis expands along with the whole
surface.}
 \end{center}
 \end{figure}

Now let us recall what was said in the paragraph containing (\ref{3.4}): $R(t,
z)$ becomes zero {\it only} at $t = t_B$. At any $t > t_B$, $R > 0$ for all $z$,
even at $M = 0$ as (\ref{3.6}) shows. Thus, the surfaces in Fig.
\ref{multisheet} can extend down to the axis $R = 0$ only if, at the given
instant $t = t_1$, the function $t_B(z)$ attains the value $t_1$ at some $z =
z_1$: then $t = t_B$ at $z = z_1$, so $R(t_1, z_1) = 0$. This is illustrated in
Fig. \ref{holenohole}. If $t_2 > t_B(z)$ at all $z$, then $R(t_2, z)$ is nowhere
zero. Let $R_0 > 0$ be the smallest lower bound of $R(t_2,z)$; then $R(t_2,z)
\geq R_0 > 0$ at all $z$, and the surface shown in Fig. \ref{multisheet} has a
hole of radius $R_0$ around the axis. Since $R,_t > 0$, $R_0$ is an increasing
function of $t$, and the radius of the hole increases with $t$. This is
illustrated in Fig. \ref{multisheetexp}.

So far, we have considered $R$ as an independent variable within the space $t =
t_0$. Since it is a function of $z$, the parameter along the radial direction in
Figs. \ref{multisheet} and \ref{multisheetexp} is in fact the coordinate $z$.
Now let us recall that $R$ is also a function of $t$ and that at every $z$ there
exists such a $t$ ($t = t_B$), at which $R = 0$. Thus, as we consider the spaces
$t = t_0$ at consecutive values of $t_0$, the surfaces depicted in those figures
get gradually ``unglued'' from the Big Bang set (which is represented by the
axis of symmetry $R = 0$), and expand sideways. At the moment, at which $t_0$
begins to obey $t_0 > t_B$ for all $z$, the surface becomes completely detached
from the axis and continues to expand sideways. This is when the hole mentioned
above first appears.

\begin{figure}[h]
 \begin{center}
  \includegraphics[scale=0.65]{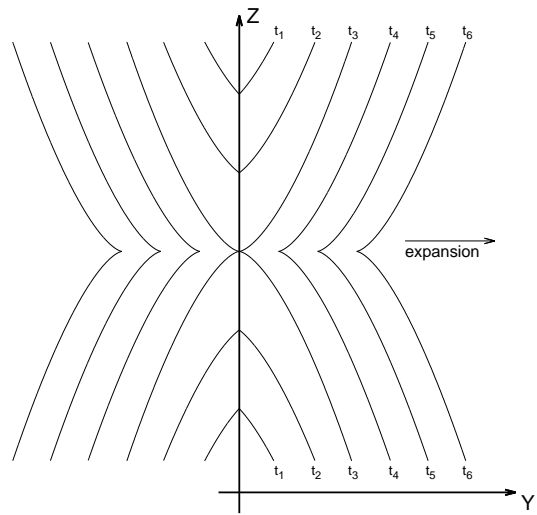}
 \caption{
 \label{evolwithedge}
 \footnotesize
Evolution of the surface from the lower panel of Fig. \ref{multisheet}. The
figure shows the axial cross-section of the surface at several time instants,
$t_1 < \dots < t_6$. The Big Bang goes off along the $Z$ axis, beginning at the
top and at the bottom, and progressing toward the middle. The instant $t_3$
corresponds to the last moment when the surface has no hole. Multiple covering
not shown.}
 \end{center}
 \end{figure}

Let us also note the double sign in the definition of $Z$, (\ref{6.19}), which
was not taken into account in Figs. \ref{multisheet} and \ref{multisheetexp}. It
means that each of those surfaces has its mirror-image attached at the bottom.
In summary, the evolution of those surfaces progresses as shown in Fig.
\ref{evolwithedge}. The functions used for this picture are $M = 10 |z|^3$, $E =
0.6 + 0.5 {\rm e}^{-|z|}$, $t_B = -10^3 |z| + 100$. The time instants are
$\left(t_1, \dots, t_6\right) = (1, 50, 100, 300, 500, 700)$.

The nondifferentiable cusp at the plane of symmetry is a consequence of the
assumption $R,_z > 0$: to avoid a singularity in the metric (\ref{5.1}), $E >
1/2$ must hold everywhere, and then (\ref{6.19}) implies $|\dril Z R| < \infty$
everywhere. This means that the upper half of the surface cannot go over
smoothly into the lower half.

\begin{figure}[h]
 \begin{center}
  \includegraphics[scale=0.65]{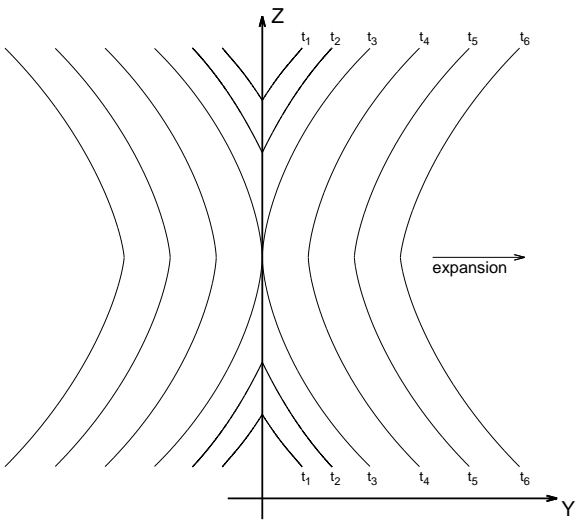}
 \caption{
 \label{evolwithnoedge}
 \footnotesize
The analogue of Fig. \ref{evolwithedge} for the situation when $R,_z = 0$ at
some $z = z_n$ (in the middle horizontal plane). Then the upper half of each
constant-$(t,\varphi)$ surface goes over smoothly into the lower half. }
 \end{center}
 \end{figure}

Let us now consider the case when $R,_z = 0$ at some $z = z_n$. To prevent a
shell crossing at $z_n$, $E(z_n) = 1/2$ must also hold, so that $\lim_{z \to
z_n}\left(R,_z/\sqrt{2E - 1}\right)$ is finite. This implies that
$\left.R,_z\right|_{z_n} = 0$ for all $t$ (i.e., that the extremum of $R$ is
comoving), and then $M,_z = E,_z = 0$ at $z = z_n$ from (\ref{2.9}). This is an
analogue of a neck -- an entity well known from studies of the Lema\^{\i}tre --
Tolman model \cite{PlKr2006}. But, as remarked under (\ref{6.20}), we have
$\dril Z R \to \pm \infty$ where $E \to 1/2$. The evolution then looks like in
Fig. \ref{evolwithnoedge}. The functions used for drawing it are $M = 10^2
|z|^3$, $E = 0.5 + |z|^{3/2}$, $t_B = -10^3 |z|^2 + 100$, and the time instants
are $\left(t_1, \dots, t_6\right) = (1, 50, 100, 200, 300, 400)$.

The minimum of $R$ with respect to $z$ need not exist in any space of constant
$t$. (By minimum we mean not only a differentiable minimum similar to the one in
Fig. \ref{evolwithnoedge}, but also a cusp at the minimal value like the one in
Fig. \ref{evolwithedge}.) It will not exist when the function $t_B(z)$ has no
upper bound, i.e., when the Big Bang keeps going off forever, moving to ever new
locations. In that case, the surfaces shown in the upper half of Fig.
\ref{evolwithedge} will never get detached from the Big Bang set, only the
vertex of each conical surface will keep proceeding along the axis. In the
special case $t_B =$ constant, the whole surface of constant $t$ and $\varphi$
gets ``unglued'' from the axis $R = 0$ at the same instant. The image would look
similar to Fig. \ref{evolwithnoedge}, but there would be no conical surfaces
with the vertices progressing along $R = 0$. The generators of the surface in
the picture are in general not vertical. They become vertical when $R,_z = 0$
everywhere, i.e., when $R= R(t)$, which can happen only in the $\beta,_z = 0$
family of Szekeres solutions that we do not discuss here.

\section{Spaces of constant $t$ in the general quasi-hyperbolic
case}\label{diffshape}

\setcounter{equation}{0}

\begin{figure}
% \begin{center}
 ${}$ \\[-10mm]
 \hspace{-30mm}
  \includegraphics[scale=0.5]{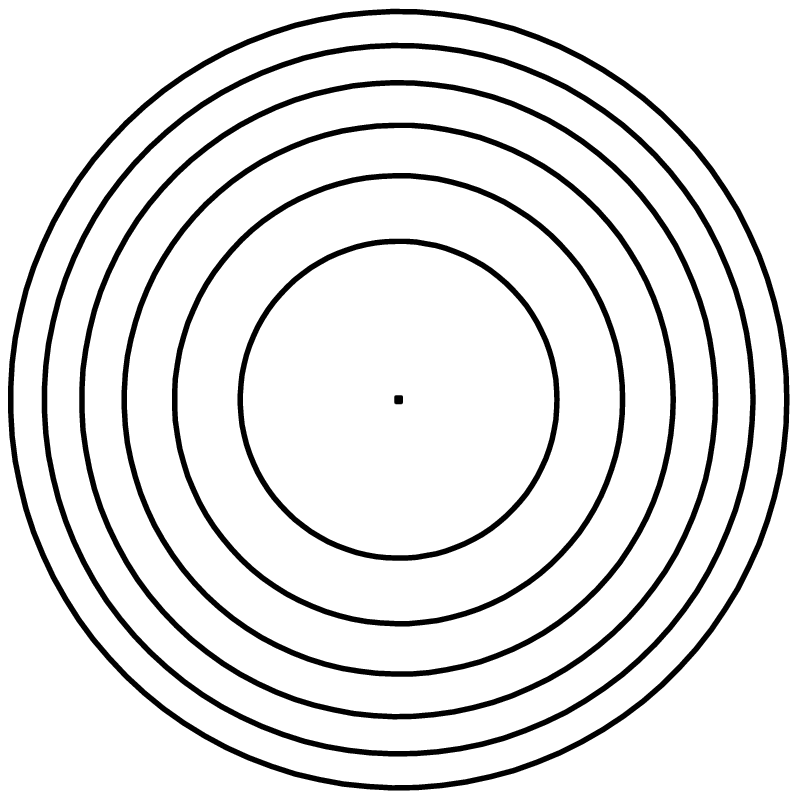}
 \hspace{-10mm}
% ${}$ \\[-10mm]
  \includegraphics[scale=0.25]{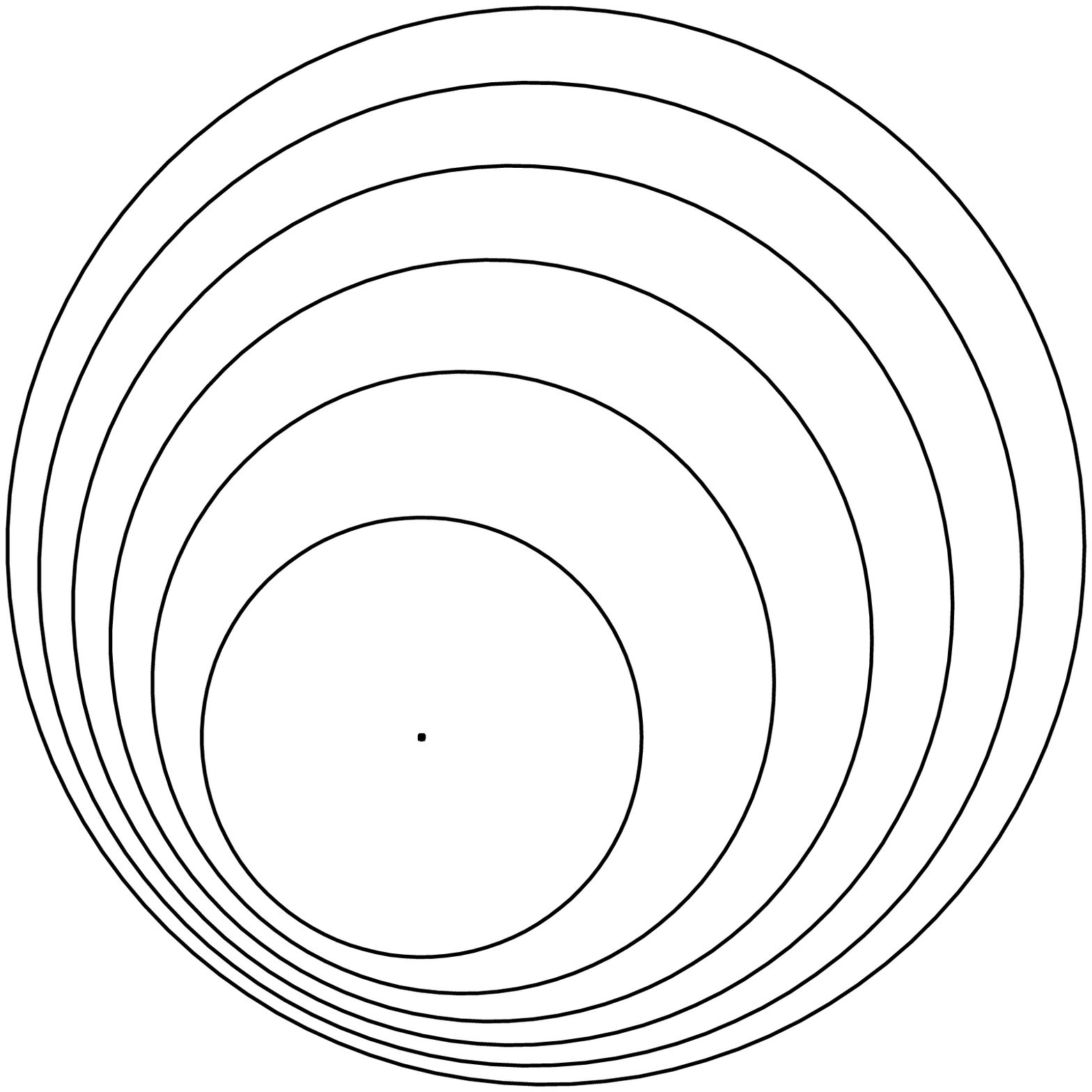}
 \caption{
 \label{concentric}
 \footnotesize
{\bf Left:} The view from above of the surface from the lower panel in Fig.
\ref{multisheet}. The circles are images of the curves of constant $R$ (and thus
of constant $z$). {\bf Right:} An example of a corresponding image in the
general quasi-hyperbolic case. Now the geodesic distance between the circles
depends on the position along the circle. }
% \end{center}
 \end{figure}

The main difference between the hyperbolically symmetric case, where ${\cal
E},_z = 0$, and the full quasi-hyperbolic case, where ${\cal E},_z \neq 0$, is
seen in (\ref{2.8}). Consider two surfaces $S_1$ and $S_2$ such that $t = t_0 =$
constant on both, $z = z_1$ on $S_1$ and $z = z_2$ on $S_2$. When ${\cal E},_z =
0$, the geodesic distance between $S_1$ and $S_2$ along a curve of constant $(x,
y)$ is the same for any $(x, y)$. When ${\cal E},_z \neq 0$, this distance
depends on $(x, y)$ and varies as the functions $P(z)$, $Q(z)$ and $S(z)$
dictate. Figure \ref{concentric}, left panel, shows an exemplary family of
constant-$R$ curves in a single surface of constant $t$ and $\varphi$ with
${\cal E},_z = 0$. This is a surface of constant $t$ and $(y/x)$ in the
coordinates of (\ref{2.8}) -- a contour map of the surface from the lower panel
in Fig. \ref{multisheet}. With a general ${\cal E}(x, y, z)$ the geodesic
distance between the constant-$R$ curves will depend on the position along each
curve, and the whole family would look like in the lower panel of Fig.
\ref{concentric}. \footnote{Fig. \ref{concentric} is in fact deceiving. The
curves shown there as circles are images of infinite curves, as explained under
(\ref{6.5}), and each image is covered an infinite number of times.}

\section{Interpretation of the mass functions $M(z)$ and ${\cal
M}(z)$ in the quasi-spherical case}\label{sphmassinterpr}

\setcounter{equation}{0}

In the quasi-spherical case, the function $M(z)$ of (\ref{2.9}), by analogy with
the Newtonian and the Lema\^{\i}tre -- Tolman cases, is understood as the active
gravitational mass inside the sphere of coordinate radius $z$. In fact, it is
puzzling why it depends only on $z$ when the mass density (\ref{2.10}) so
prominently depends also on $t$, $x$ and $y$. Somewhat miraculously, as shown
below, the denominator in (\ref{2.10}) is canceled by the $\sqrt{- g_{11}}$ term
inside the integral $\int \rho \sqrt{\left|g_3\right|} {\rm d}_3 x$ that
determines the mass in a sphere.\footnote{$g_3$ is the determinant of the metric
of the 3-space $t =$ constant in (\ref{2.1}).} The term containing ${\cal E},_z$
in the numerator gives a zero contribution to the integral. This is consistent
with the fact, known from electrodynamics, that the total charge of a dipole is
zero (see Refs. \cite{DeSo1985,PlKr2006,KrBo2012} for the splitting of
(\ref{2.10}) into the monopole and the dipole part in the quasi-spherical case).
It is also consistent with the result of Bonnor \cite{Bonn1976a, Bonn1976b} that
the Szekeres solution can be matched to the Schwarzschild solution.

The considerations of this and the next three sections are intended to prepare
the ground for an analogous investigation in the quasi-hyperbolic case further
on. The questions we seek to answer are: Can $M$ still be interpreted as mass,
and where does the mass $M(z)$ reside when a surface of constant $z$ has
infinite surface area?

Let us calculate, in the quasi-spherical case, the amount of rest mass within
the sphere of coordinate radius $z$ at coordinate time $t$, assuming that $z =
z_0$ is the center, where the sphere has zero geometrical radius (see Ref.
\cite{HeKr2002}). This amount equals ${\cal M} = \int_{\cal V} \rho
\sqrt{\left|g_3\right|} {\rm d}_3 x$, where ${\cal V}$ is the volume of the
sphere and $\rho$ is the mass density given by (\ref{2.10}). Substituting for
$\rho$ and $g_3$ we get
\begin{eqnarray}
{\cal M} &=& \frac 1 {4 \pi} \int_{- \infty}^{+ \infty} {\rm d} x \int_{-
\infty}^{+ \infty} {\rm d} y \int_{z_0}^z {\rm d} u \nonumber \\
&& \left[\frac {M,_u(u)} {\sqrt{1 + 2E} {\cal E}^2} - \frac {3M {\cal E},_u}
{\sqrt{1 + 2E} {\cal E}^3}\right], \label{8.1}
\end{eqnarray}
where $u$ is the running value of $z$ under the integral. Note that ${\cal E}$
is the only quantity that depends on $x$ and $y$, it is an explicitly given
function, and so the integration over $x$ and $y$ can be carried out:
\begin{equation}\label{8.2}
\int_{- \infty}^{+ \infty} {\rm d} x \int_{- \infty}^{+ \infty} {\rm d} y \frac
1 {{\cal E}^2} = 4 \pi, \quad \int_{- \infty}^{+ \infty} {\rm d} x \int_{-
\infty}^{+ \infty} {\rm d} y \frac {{\cal E},_z} {{\cal E}^3} = 0.
\end{equation}
(The first of these just confirms that this is the surface area of a unit
sphere.) Using this in (\ref{8.1}) we get
\begin{equation}\label{8.3}
{\cal M} = \int_{z_0}^z \frac {M,_u} {\sqrt{1 + 2E}}(u) {\rm d} u,
\end{equation}
which is the same relation as in the LT model,\footnote{For the quasi-spherical
case also other integral relations are similar to the ones in the L--T model,
which is a consequence of the fact that the dipole contribution vanishes after
averaging over $x$-$y$ surface \cite{Bole2009}.} and shows that $1 / \sqrt{1 +
2E}$ is the relativistic energy defect/excess function (when $2E < 0$ and $2E
>0$ respectively).

In the quasi-spherical case we are able to calculate the integral with respect
to $x$ and $y$ over the whole $(x, y)$ surface because its surface area is
finite. Such a calculation cannot be repeated for the $\varepsilon \leq 0$ cases
because both integrals analogous to (\ref{8.2}) are infinite. Let us then
consider what happens with $M$ and ${\cal M}$ when we calculate the integrals in
(\ref{8.2}) over a part of the sphere.

\section{The mass in the spherically symmetric case}\label{sphermass}

\setcounter{equation}{0}

In nearly all the papers concerning the LT model and the quasi-spherical
Szekeres model it was assumed that each space of constant $t$ has its center of
symmetry (in the LT case) or \textit{origin} (in the quasi-spherical Szekeres
case), where $M = 0$ and $R = 0$ at all times. We will assume the same here, but
\textit{this is an assumption}. It is possible that the center of symmetry is
not within the spacetime in the LT case, and the corresponding quasi-spherical
Szekeres generalization will then have no ``origin''.\footnote{The authors are
aware of just one paper, in which LT models without a center of symmetry were
considered. These are the ``in one ear and out the other'' and the ``string of
beads'' models of Hellaby \cite{Hell1987}, described also in Ref.
\cite{PlKr2006}. In both of them, $M \neq 0$ throughout the space.}

For the beginning we will consider the spherically symmetric
(Lema\^{\i}tre--Tolman) subcase, in which $P,_z = Q,_z = S,_z = 0$, so $P = Q =
0$ can be achieved by coordinate transformations. Then ${\cal E},_z \equiv 0$ in
(\ref{8.2}). Now suppose that we calculate the integral in the first of
(\ref{8.2}) over a circular patch $C$ of the sphere (circular in order that no
$(x, y)$ dependence appears from the boundary shape). The boundary of $C$ is an
intersection of the sphere with a cone whose vertex is at the center of the
sphere. Let the vertex angle $\theta$ of the cone be $\pi/n$. This translates to
the radius of $C$ in the original $(x, y)$ coordinates being $u_0 = S \tan
(\pi/2 n) \df S \beta$. Then we have in place of the first of (\ref{8.2})
\begin{eqnarray}\label{9.1}
\int_C {\rm d}_2xy \frac 1 {{\cal E}^2} &=& 4 \pi \frac {{u_0}^2} {S^2 +
{u_0}^2} \equiv 2 \pi \left[1 - \cos (\pi/n)\right] \nonumber \\
&\equiv& 4 \pi \frac {\beta^2} {1 + \beta^2}.
\end{eqnarray}
This tends to $4 \pi$ when $n \to 1$ ($u_0 \to \infty$). Note that the final
result does not depend on $S$ -- this happened because we have chosen the
coordinate radius in each circle, $\sqrt{x^2 + y^2} = u_0$, to be a fixed
multiple of $S$.

In the spherically symmetric case now considered, we choose the same cone to
define the circles of integration in (\ref{9.1}) in all surfaces of constant
$z$. Instead of (\ref{8.3}) we get for the amount of rest mass within the cone,
${\cal M}_C$:
\begin{eqnarray}\label{9.2}
{\cal M}_C &=& \frac 1 {4 \pi} \int_C {\rm d}_2xy \int_{z_0}^z {\rm d} u \frac
{M,_u} {\sqrt{1 + 2E} {\cal E}^2} \nonumber \\
&=& \frac {\beta^2} {1 + \beta^2} \int_{z_0}^z \frac {M,_u} {\sqrt{1 + 2E}}(u)
{\rm d} u.
\end{eqnarray}
The term of (\ref{8.1}) that contained ${\cal E},_u$ disappeared here in
consequence of the assumed spherical symmetry, but it will not disappear when we
go over to the general case, and its contribution will have to be interpreted.

\section{Symmetry transformations of a sphere in the coordinates of
(\ref{2.12})}\label{symtransf}

\setcounter{equation}{0}

In the spherically symmetric case rotations around a point are symmetries of the
space. In this case, if we rotate the whole cone around its vertex to any other
position, eq. (\ref{9.2}) will not change. Each $z =$ const circle will then be
rotated by the same angles to its new position, and the result of such a
rotation will be a cone isometric to the original one.

However, in the general, nonsymmetric case the spheres $z =$ constant are not
concentric. Suppose we build, in the general case, a surface composed of
circles, each circle taken from a different sphere. If we rotate each sphere by
the same angles, whatever surface existed initially, will be deformed into a
shape non-isometric to the original one. We want to calculate the effect of such
a transformation, and for this purpose we need the formulae for the $O(3)$
rotations in the $(x, y)$ coordinates. We calculate them now.

The generators of spherical symmetry, in the ordinary spherical coordinates, are
\cite{PlKr2006}
\begin{eqnarray}\label{10.1}
J_1 &=& \pdr {} {\varphi}, \qquad J_2 = \sin \varphi \pdr {} {\vartheta} + \cos
\varphi \cot \vartheta \pdr {} {\varphi}, \nonumber \\
J_3 &=& \cos \varphi \pdr {} {\vartheta} - \sin \varphi \cot \vartheta \pdr {}
{\varphi}
\end{eqnarray}
We transform these to the $(x, y)$ coordinates of (\ref{2.12}), for the
beginning with $P = Q = 0, S = 1$, by
\begin{equation}\label{10.2}
x = \cot (\vartheta/2) \cos(\varphi), \qquad y = \cot (\vartheta/2)
\sin(\varphi)
\end{equation}
and obtain
\begin{eqnarray}\label{10.3}
J_1 &=& x \pdr {} y - y \pdr {} x, \nonumber \\
J_2 &=& 2xy \pdr {} x + \left(1 - x^2 + y^2\right) \pdr {} y, \nonumber \\
J_3 &=& \left(1 + x^2 - y^2\right) \pdr {} x + 2xy \pdr {} y.
\end{eqnarray}
The transformations generated by $J_1$ are rotations in the $(x, y)$ plane. To
find the transformations generated by $J_3$ we have to solve (see Ref.
\cite{PlKr2006} for explanations):
\begin{equation}\label{10.4}
\dr {x'} {\lambda} = 1 + {x'}^2 - {y'}^2, \qquad \dr {y'} {\lambda} = 2x'y',
\end{equation}
where $\lambda$ is the parameter of the group generated by $J_3$. The general
solution of this is
\begin{eqnarray}\label{10.5}
y' &=& 1 / U_1, \qquad U_1 \df C + \sqrt{C^2 - 1} \cos(2\lambda + D), \nonumber
\\
x' &=& \left[\sqrt{C^2 - 1} \sin(2 \lambda + D)\right] / U_1,
\end{eqnarray}
where $C$ and $D$ are arbitrary constants of integration. These have to obey the
initial conditions $\left.(x', y')\right|_{\lambda = 0} = (x, y)$. After solving
for $C$ and $D$ this leads to
\begin{eqnarray}\label{10.6}
&& x' = \left[2x \cos (2 \lambda) + \left(1 - x^2 - y^2\right) \sin (2
\lambda)\right] / U_3 \nonumber \\
&& U_3 \df 1 + x^2 + y^2 + \left(1 - x^2 - y^2\right) \cos (2 \lambda) \nonumber
\\
&& \ \ \ \ \ \ \ \ \ \ \ - 2x \sin (2 \lambda), \nonumber \\
&& y' = 2y / U_3.
\end{eqnarray}

It is instructive to calculate the effect of the transformation (\ref{10.6}) in
the $(\vartheta, \varphi)$ coordinates of (\ref{10.2}). Let us then take a point
of coordinates $(x, y) = (x, 0)$, i.e., $(\vartheta, \varphi) = (\vartheta, 0)$,
and let us apply (\ref{10.6}) to it. After a little trigonometry we get
\begin{eqnarray}\label{10.7}
\tan (\vartheta'/2) &=& \frac {1 - \cos \vartheta \cos (2\lambda) - \sin
\vartheta \sin (2\lambda)} {\sin \vartheta \cos (2 \lambda) - \cos \vartheta
\sin (2 \lambda)} \nonumber \\
&\equiv& \tan (\vartheta / 2 - \lambda) \Longrightarrow \vartheta' = \vartheta -
2 \lambda,
\end{eqnarray}
i.e., (\ref{10.6}) is equivalent to rotating the sphere around the $(\vartheta,
\varphi) = (\pi / 2, \pi / 2)$ axis by the angle $(- 2 \lambda)$.

It can now be verified that the quantity:
\begin{equation}\label{10.8}
I(x, y) \df \frac {1 + x^2 + y^2} {2y} \equiv \frac {1 + {x'}^2 + {y'}^2} {2y'}
\end{equation}
is an invariant of the transformation (\ref{10.6}). The set $I = C$ is the
circle $x^2 + (y - C)^2 = C^2 - 1$.

An arbitrary circle of radius $A$ and center at $x = y = 0$, $x^2 + y^2 = A^2$,
is transformed by (\ref{10.6}) into the circle
\begin{eqnarray}\label{10.9}
&& \left[x' - \frac {\left(1 + A^2\right) \sin (2 \lambda)} {\left(1 +
A^2\right) \cos (2 \lambda) + 1 - A^2}\right]^2 + {y'}^2 \nonumber \\
&& = \frac {4A^2} {\left[\left(1 + A^2\right) \cos (2 \lambda) + 1 -
A^2\right]^2}.
\end{eqnarray}
The coordinate radius of the $(x', y')$ circle is different from the original
radius $A$ except when $\lambda  = 0$, which is the identity transformation. But
the coordinate radius is not an invariantly defined quantity. An invariant
measure of the circle, its surface area, does not change under the
transformation (\ref{10.6}), and neither does the invariant distance between any
two points, as we show below.

The Jacobian of the transformation (\ref{10.6}) is
\begin{equation}\label{10.10}
\pdr {(x', y')} {(x, y)} = \frac 4 {{U_3}^2}.
\end{equation}
Together with (\ref{10.8}) and (\ref{10.6}) this shows that the surface element
under the integral (\ref{9.2}), $4 {\rm d} x {\rm d} y / {\cal E}^2$, does not
change in form after the transformation. (This must be so, since (\ref{10.6}) is
just a change of variables that does not change the value of the integral.)
Thus, (\ref{10.6}) preserves the area of any circle, as is appropriate for a
symmetry.

The invariant distance between the points $(x, y) = (0, 0)$ and $(x, y) = (A,
0)$ (i.e., the invariant radius of the original circle referred to in
(\ref{10.9})) is, from (\ref{2.7}) -- (\ref{2.8}) with $\varepsilon = +1, P = Q
= 0, S = 1$:
\begin{equation}\label{10.11}
\int_0^A \frac {{\rm d} x} {1 + x^2} = 2 \arctan A.
\end{equation}
The image under (\ref{10.6}) of any point $(x, 0)$ is $(x_1(x), 0)$, where, from
(\ref{10.6}):
\begin{equation}\label{10.12}
x_1(x) = \frac {2x \cos (2 \lambda) + \left(1 - x^2\right) \sin (2 \lambda)} {1
+ x^2 + \left(1 - x^2\right) \cos (2 \lambda) - 2x \sin (2 \lambda)}.
\end{equation}
Thus, the image of $(0, 0)$ is $(x_0, 0)$, where
\begin{equation}\label{10.13}
x_0 = \frac {\sin (2 \lambda)} {1 + \cos (2 \lambda)}.
\end{equation}
The invariant distance between the images of $(0, 0)$ and of $(A, 0)$ is then
\begin{equation}\label{10.14}
\int_{x_1(0)}^{x_1(A)} \frac {{\rm d} x_1} {1 + {x_1}^2} = 2 \left.\arctan
(x_1)\right|_{x_0}^{x_1(A)} \equiv 2 \arctan A,
\end{equation}
by employing the identity $\arctan \alpha - \arctan \beta = \arctan
\left[(\alpha - \beta) / (1 + \alpha \beta)\right]$. This certifies that the
invariant distance between the center of a circle and a point on the circle is
the same as the invariant distance between their images (but the image of the
center is no longer the center of the image-circle, compare (\ref{10.9}) and
(\ref{10.13})).

The transformations generated by $J_2$ result from those for $J_3$ by
interchanging $x'$ with $y'$ and $x$ with $y$; then all the conclusions about
invariant properties follow also for these transformations, and, in consequence,
for any composition of (\ref{10.6}) with them.

\vspace{-5mm}

\section{The mass in the general quasi-spherical case}\label{genqsphermass}

\setcounter{equation}{0}

\vspace{-4mm}

Now let us consider the general case, and integrals analogous to (\ref{8.2}),
where the $(x, y)$ integration extends only over a circular subset of each
sphere, the radius of each circle being a fixed multiple of $S$. In the general
case, each sphere has a geometrically preferred center at $(x, y) = (P(z),
Q(z))$, and, for the beginning, we choose the center of the disc of integration
$C$ at that point. As before, the radius of each circle will be a fixed multiple
of $S$: $\sqrt{x^2 + y^2} \df u_0 = S \beta$. This means, this time the volume
of integration will not be a simple cone, but a `wiggly cone' -- the circles in
the different $z =$ const surfaces will have their centers not on a straight
line orthogonal to their planes, but on the curve given by the parametric
equations $x = P(z)$, $y = Q(z)$ that is not orthogonal to the planes of the
circles. The result (\ref{9.1}) still holds within each $z =$ const surface, but
the analogue of the second integral in (\ref{8.1}), calculated over the interior
of the `wiggly cone' here, will no longer be zero. Instead, introducing in each
$z =$ const surface the polar coordinates $x = P + u \cos \varphi$, $y = Q + u
\sin \varphi$, we get
 \begin{widetext}
\begin{eqnarray}\label{11.1}
&& \int_C {\rm d}_2xy \frac {{\cal E},_z} {{\cal E}^3} = \int_0^{2 \pi} {\rm d}
\varphi \int_0^{u_0} u {\rm d} u  \frac {(S,_z/2) \left(1 - u^2 / S^2\right) -
\frac 1 S \left(P,_z u \cos \varphi + Q,_z u \sin \varphi\right)} {\left(S^3 /
8\right) \left(1 + u^2 / S^2\right)^3} \nonumber \\
&& \equiv 8 \pi SS,_z \int_0^{u_0} \frac {uS,_z \left(S^2 - u^2\right)}
{\left(S^2 + u^2\right)^3} {\rm d} u = 4 \pi SS,_z \frac {{u_0}^2} {\left(S^2 +
{u_0}^2\right)^2} = 4 \pi \frac {S,_z} S \frac {\beta^2} {\left(1 +
\beta^2\right)^2}.
\end{eqnarray}
In agreement with (\ref{8.2}) this goes to zero when $u_0 \to \infty$.
Consequently, from (\ref{8.1}), (\ref{9.2}) and (\ref{11.1}), the total mass
within the wiggly cone is
\begin{equation}\label{11.2}
{\cal M} = \frac {\beta^2} {1 + \beta^2} \int_{z_0}^z \frac {M,_u} {\sqrt{1 +
2E}}(u) {\rm d} u - 3 \frac {\beta^2} {\left(1 + \beta^2\right)^2} \int_{z_0}^z
\frac {MS,_u} {S \sqrt{1 + 2E}}(u) {\rm d} u.
\end{equation}
 \end{widetext}
It contains a contribution from $S,_z$ that is decreasing with increasing
$\beta$, i.e., the greater volume we take, the less significant the contribution
from $S,_z$ gets. It will vanish when the integrals extend over the whole
infinite range of $x$ and $y$ (in the limit $\beta \to \infty$). This can be
interpreted so that in a wiggly cone the dipole components of mass distribution
do contribute to ${\cal M}$ -- but less and less as the volume of the cone
increases. Thus, with such choice of the integration volume $M$ does not have an
immediate interpretation -- but it becomes proportional to the mass within the
cone in the spherically symmetric limit.

\begin{figure}[h]
 \begin{center}
 ${}$ \\[-5mm]
 \includegraphics[scale=0.5]{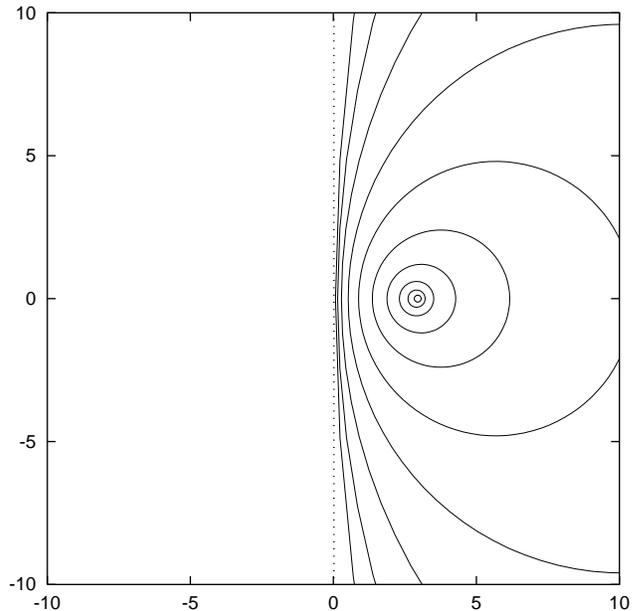}
 \caption{
 \label{circfamily}
 \footnotesize
An axial cross-section through the family of spheres given by (\ref{11.3}). }
 \end{center}
 \end{figure}

\begin{figure*}
 \begin{center}
 ${}$ \\[-5mm]
 \hspace{-90mm}
 \includegraphics[scale=0.38]{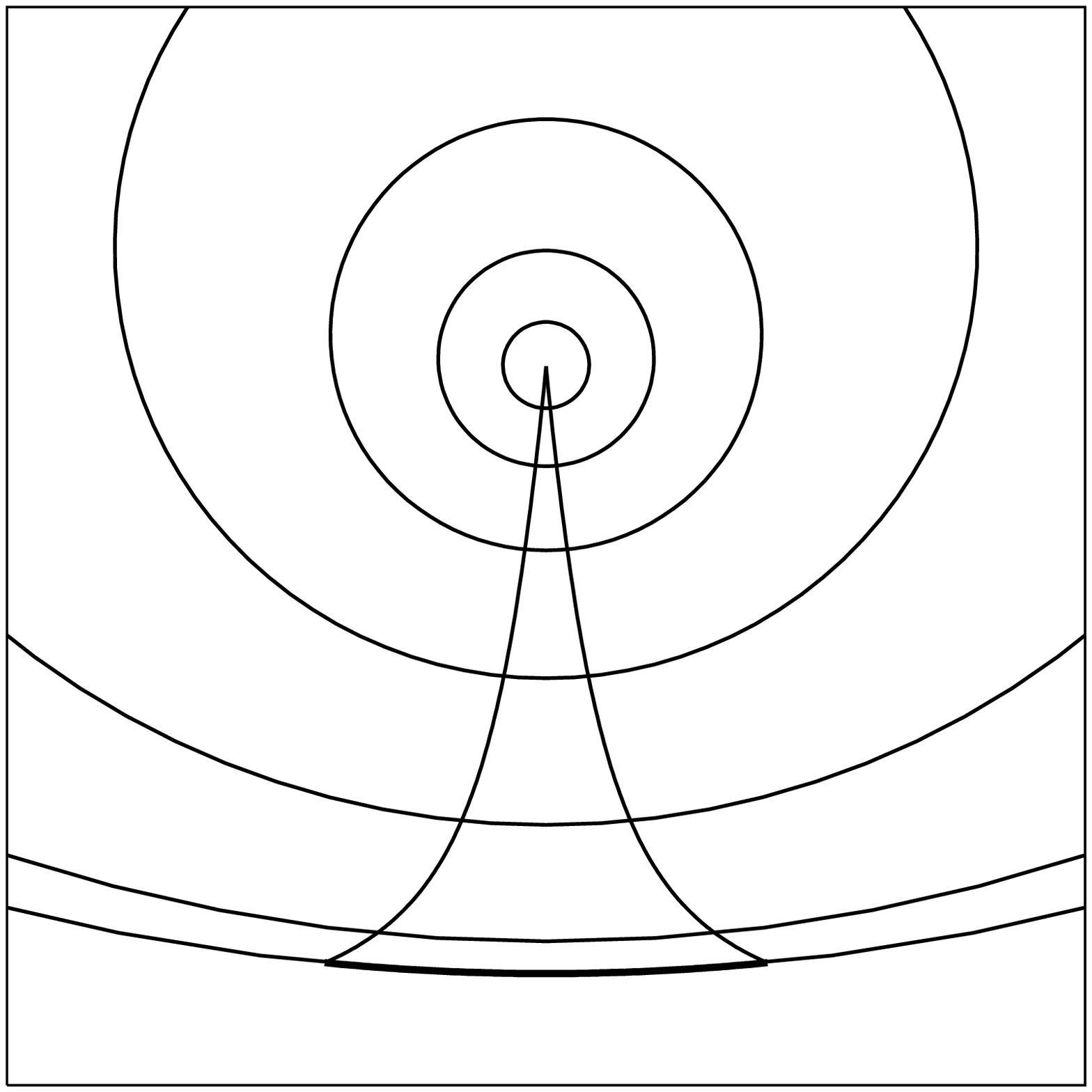}
 ${}$ \\[-62mm]
 \hspace*{65mm}
  \includegraphics[scale=0.5]{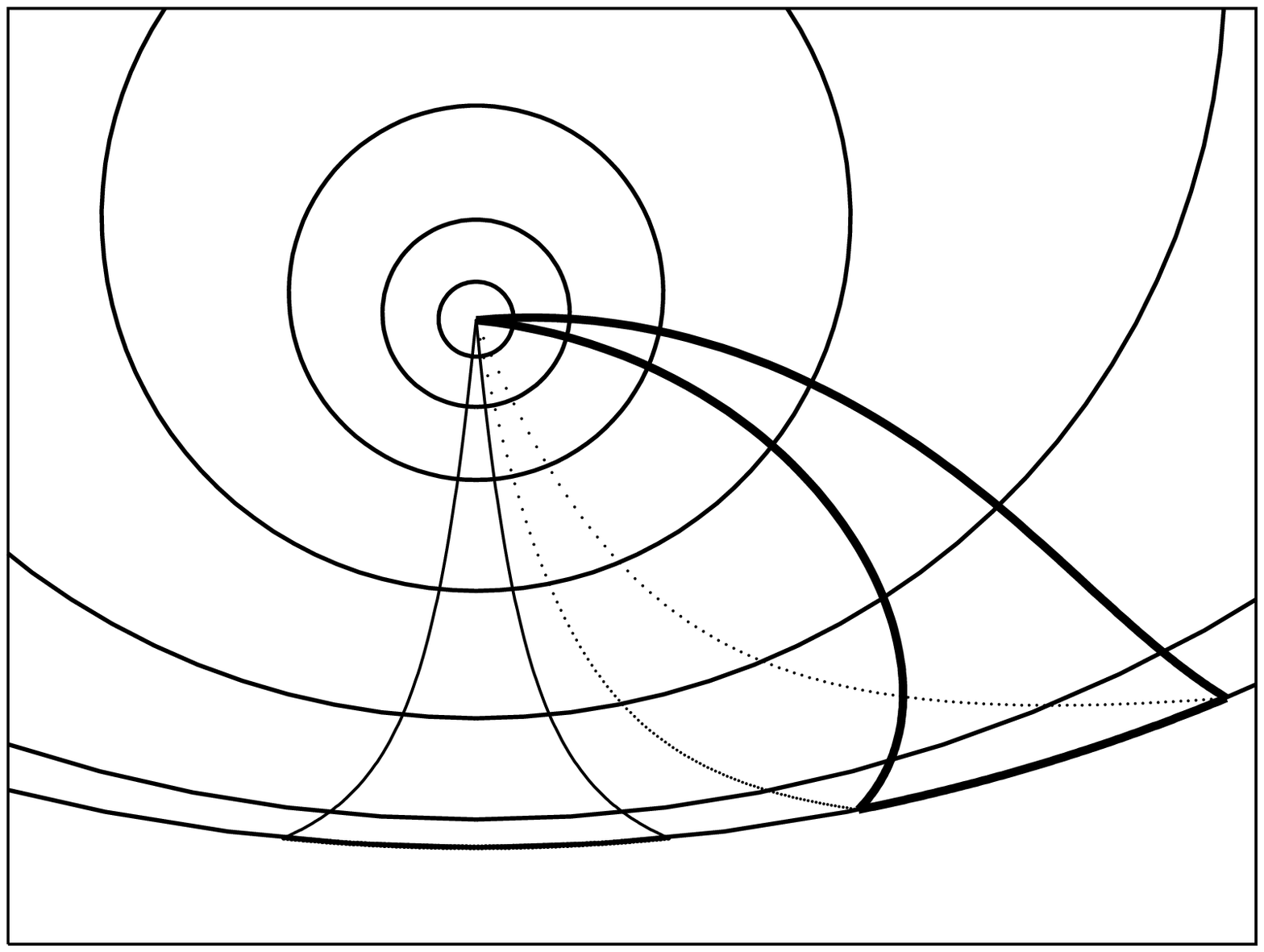}
 \caption{
 \label{wigcones}
 \footnotesize
{\bf Left graph:} a wiggly cone constructed for the family of spheres shown in
Fig. \ref{circfamily}. The vertex angle for this cone is $\pi/32$. {\bf Right
graph:} the result of the transformation (\ref{11.4}) applied to the cone from
the left graph. The initial cone is shown in thin lines. The rest mass contained
in the new wiggly cone is the same as it was in the initial cone. Dotted lines
show the image of the initial cone that would result if each circle of
intersection of the original cone with a sphere were rotated by the same angle
around the center of its sphere. }
 \end{center}
 \end{figure*}

As an example, consider the axially symmetric family of spheres whose axial
cross-section is shown in Fig. \ref{circfamily}. The circles are given by the
equation
\begin{equation}\label{11.3}
\left(x - \sqrt{b^2 + u^2}\right)^2 + y^2 = u^2 ~, \\
\end{equation}
where $b$ is a constant that determines the center of the limiting circle of
zero radius, while $u$ is the radius of the circles. (The same family of spheres
was used in Ref. \cite{HeKr2008} to construct Szekeres coordinates for a flat
space.) Figure \ref{wigcones}, left graph, shows the initial wiggly cone
constructed for these spheres -- the one referred to in (\ref{11.1}) and
(\ref{11.2}).

We derived the transformation (\ref{10.6}) in the coordinates in which the
constants $(P, Q, S)$ were set to $(0, 0, 1)$ by coordinate transformations.
With general values of $(P, Q, S)$, the result would be
 \begin{widetext}
\begin{eqnarray}\label{11.4}
\frac {x' - P} S &=& \left[2 \frac {x - P} S \cos (2 \lambda) + \left(1 - \frac
{(x - P)^2 + (y - Q)^2} {S^2}\right) \sin (2
\lambda)\right] / U_4 \nonumber \\
y' - Q &=& 2 \frac {y - Q} {U_4}, \nonumber \\
U_4 &\df& 1 + \frac {(x - P)^2 + (y - Q)^2} {S^2} + \left[1 - \frac {(x - P)^2 +
(y - Q)^2} {S^2}\right] \cos (2 \lambda) - 2 \frac {x - P} S \sin (2 \lambda).
\end{eqnarray}
 \end{widetext}

Now let $z = z_1$ correspond to the base of the wiggly cone, where the values of
the arbitrary functions are $P_1 = P(z_1)$, $Q_1 = Q(z_1)$ and $S_1 = S(z_1)$.
Apply the transformation (\ref{11.4}) with $(P, Q, S) = (P_1, Q_1, S_1)$ to each
sphere intersecting the wiggly cone. In the base $z = z_1$ this will be a
symmetry, in other spheres this will not be a symmetry. One should in principle
calculate the effect of this transformation in other spheres on the integrands
in (\ref{9.2}) and (\ref{11.1}) to see what happens. But then, (\ref{11.4}) is
merely a change of variables under the integral that does not change the value
of the calculated integral. Thus, eq. (\ref{11.2}) applies independently of
where we choose the base of the wiggly cone, and its position that we chose
initially (each circle had its center in the geometrically distinguished center
of the $(x, y)$ surface) is only necessary to fix the relation between circles
corresponding to different values of $z$.

The result of the transformation (\ref{11.4}) applied to the wiggly cone of the
left graph in Fig. \ref{wigcones} is shown in the same figure in the right
graph.

\section{The symmetry transformations in the hyperbolically symmetric
case}\label{hypsym}

\setcounter{equation}{0}

Before proceeding to the case of interest, let us consider the hyperbolically
symmetric subcase, in which $P,_z = Q,_z = S,_z = 0$, and so $P = Q = 0$ by a
transformation of $x$ and $y$. The symmetry group of the resulting metric is a
subgroup of that for the corresponding vacuum solution \cite{HeKr2008}:
\begin{eqnarray}\label{12.1}
{\rm d} s^2 &=& - \left(1 + \frac {2m} R\right) {\rm d} T^2 + \frac 1 {1 + 2m /
R}\ {\rm d} R^2 \nonumber \\
&-& R^2 \left({\rm d} \vartheta^2 + \sinh^2 \vartheta {\rm d} \varphi^2\right).
\end{eqnarray}
The full set of Killing vectors for this metric is
\begin{eqnarray}\label{12.2}
{k_1}^{\alpha} &=& {\delta_0}^{\alpha}, \qquad {k_4}^{\alpha} =
{\delta_3}^{\alpha}, \nonumber \\
{k_2}^{\alpha} &=& \cos \varphi {\delta_2}^{\alpha} - \coth \vartheta \sin
\varphi {\delta_3}^{\alpha}, \nonumber \\
{k_3}^{\alpha} &=& \sin \varphi {\delta_2}^{\alpha} + \coth \vartheta \cos
\varphi {\delta_3}^{\alpha},
\end{eqnarray}
where $(x^0, x^1, x^2, x^3) = (T, R, \vartheta, \varphi)$. To find the
symmetries explicitly, we have to transform (\ref{12.2}) to the coordinates of
(\ref{2.8}).

We are interested in the transformations within an $(x, y)$ surface, and those
are generated by $k_2$, $k_3$ and $k_4$. The transformation from the
$(\vartheta, \varphi)$ coordinates of (\ref{12.1}) to sheet 2 of the $(x, y)$
coordinates of (\ref{2.8}) is (\ref{5.4}). The inverse formulae are
\begin{eqnarray}\label{12.3}
\varphi &=& \arctan (y/x), \nonumber \\
\cosh \vartheta &=& \frac {1 + x^2 + y^2} {1 - \left(x^2 +
y^2\right)}, \nonumber \\
\sinh \vartheta &=& \frac {2 \sqrt {x^2 + y^2}} {1 - \left(x^2 + y^2\right)}.
\end{eqnarray}
We now transform the Killing fields (\ref{12.2}) by (\ref{12.3}). In the $(x,
y)$ coordinates we get for the generators
\begin{eqnarray}
J_1 &=& x \pdr {} y - y \pdr {} x, \label{12.4} \\
J_2 &=& \left[1 + \left(y^2 - x^2\right)\right] \pdr {} x - 2 x y \pdr {} y,
\label{12.5}\\
J_3 &=& - 2 x y \pdr {} x + \left(1 + x^2 - y^2\right) \pdr {} y.\label{12.6}
\end{eqnarray}
The $J_1$ generates rotations in the $(x, y)$ surface. To find the
transformations generated by $J_2$, we have to solve the set
\begin{equation}\label{12.7}
\dr {x'} {\lambda} = 1 + {y'}^2 - {x'}^2, \qquad \dr {y'} {\lambda} = - 2 x' y',
\end{equation}
with the initial condition that at $\lambda = 0$ we have $(x', y') = (x, y)$.
The general solution of this set is
\begin{eqnarray}\label{12.8}
y' &=& 1 / W_1, \qquad W_1 \df - C + \sqrt{C^2 + 1} \cosh (2 \lambda + D),
\nonumber \\
x' &=& (1 / W_1) \sqrt{C^2 + 1} \sinh (2 \lambda + D),
\end{eqnarray}
where $C$ and $D$ are arbitrary constants to be determined from $x'(0) = x$,
$y'(0) = y$. They are
\begin{eqnarray}\label{12.9}
\sinh D &=& 2 x / W_2, \qquad \cosh D = \left(x^2 + y^2 + 1\right) /
W_2, \nonumber \\
W_2 &\df& \sqrt{\left(x^2 + y^2 - 1\right)^2 + 4 y^2}, \nonumber \\
C &=& \frac 1 {2 y} \left(x^2 + y^2 - 1\right).
\end{eqnarray}
So, finally
\begin{eqnarray}\label{12.10}
&& x' = (1 / W_3) \left[2 x \cosh (2 \lambda) + \left(1 + x^2 + y^2\right)
\sinh (2 \lambda)\right], \nonumber \\
&& W_3 \df 1 - \left(x^2 + y^2\right) + \left(1 + x^2 + y^2\right)
\cosh (2 \lambda) \nonumber \\
&& \ \ \ \ \ \ \ \ \ + 2 x \sinh (2 \lambda), \nonumber \\
&& y' = 2 y / W_3.
\end{eqnarray}

The equations corresponding to (\ref{12.7}) for the generator $J_3$ result from
(\ref{12.7}) simply by interchanging $x'$ with $y'$. The corresponding initial
condition then results by interchanging $x$ with $y$. Thus, from (\ref{12.10})
we can read off the transformation generated by $J_3$; it is
\begin{eqnarray}\label{12.11}
&& x' = 2 x / W_4, \nonumber \\
&& W_4 \df 1 - \left(x^2 + y^2\right) + \left(1 + x^2 + y^2\right) \cosh
(2 \lambda) \nonumber \\
&& \ \ \ \ \ \ \ \ \ + 2 y \sinh (2 \lambda), \\
&& y' = (1 / W_4) \left[2 y \cosh (2 \lambda) + \left(1 + x^2 + y^2\right) \sinh
(2 \lambda)\right] \nonumber .
\end{eqnarray}

Note that the $(x', y')$ given by (\ref{12.10}) also obey the third equation in
(\ref{12.9}), so the quantity
\begin{equation}\label{12.12}
I_1 \df = \frac 1 {2 y} \left(x^2 + y^2 - 1\right)
\end{equation}
is an invariant of the transformations (\ref{12.10}). The corresponding
invariant for (\ref{12.11}) is
\begin{equation}\label{12.13}
I_2 \df = \frac 1 {2 x} \left(x^2 + y^2 - 1\right)
\end{equation}
These facts are helpful in calculations, and so is the following identity that
follows from (\ref{12.12})
\begin{equation}\label{12.14}
{x'}^2 + {y'}^2 - 1 \equiv \frac 2 {W_3} \left(x^2 + y^2 - 1\right).
\end{equation}

The fact that $I_1$ is an invariant of (\ref{12.10}) means that the
transformation (\ref{12.10}) maps the set $I_1 = C = $ constant into itself for
every $C$. This set is a circle of radius $\sqrt{C^2 + 1}$ and center in the
point $(x, y) = (0, C)$.

The inverse transformation to (\ref{12.10}) results from (\ref{12.10}) by the
substitution $\lambda \rightarrow (- \lambda)$. This can be verified using the
above identities.

These same identities can be used to verify that the transformation
(\ref{12.10}) maps the circle $x^2 + y^2 = A^2$ into the following circle in the
$(x', y')$ coordinates
\begin{eqnarray}\label{12.15}
&& \left[x' - \frac {\left(1 - A^2\right) \sinh (2 \lambda)} {1 + A^2 +
\left(1 - A^2\right) \cosh (2 \lambda)}\right]^2 + {y'}^2 \nonumber \\
&& = \frac {4 A^2} {\left[1 + A^2 + \left(1 - A^2\right) \cosh (2
\lambda)\right]^2}.
\end{eqnarray}
The radius of this new circle equals the original radius, $A$, only in two
cases: $\lambda = 0$, which is an identity transformation, or $A = 1$. In both
cases, also the center of the circle remains unchanged. The radius meant here is
a {\it coordinate radius} that has no invariant meaning. The meaningful quantity
is the geometric radius, which is the invariant distance between the center of
the circle and a point on its circumference. It can be verified that the
invariant distance between any pair of points is the same as the invariant
distance between their images.

Below we present some remarks about the transformation (\ref{12.10}). The same
statements apply to (\ref{12.11}).

The Jacobian of the transformation (\ref{12.10}) is
\begin{equation}\label{12.16}
\pdr {(x', y')} {(x, y)} = \frac 4 {{W_3}^2},
\end{equation}
which, together with (\ref{12.14}), shows that the integrand in the integral
$\int {\rm d}_2xy / {\cal E}^2$ is form-invariant under this transformation. In
this integral, (\ref{12.10}) is an ordinary change of variables, and the area of
integration in the $(x', y')$ variables will be an image of the original area
under the same transformation. This means that also the value of this integral
is an invariant of (\ref{12.10}). Thus, if we choose the region of integration
to be a circle around a point, it does not matter where the center of the circle
is because we can freely move the circle around the $(x, y)$ surface by symmetry
transformations without changing its area.

 \bigskip

\section{The mass function in the quasi-hyperbolic case}\label{quasihyp}

\setcounter{equation}{0}

In the quasi-hyperbolic case, the $(x, y)$ surfaces are infinite, so they do not
surround any finite volume. Thus, unlike in the quasi-spherical case, we should
not expect the value of the mass function $M(z)$ to correspond to a mass
contained in a well-defined volume. We should rather observe the analogy to a
solid cylinder of finite radius and infinite length in Newton's theory, in which
the mass density depends only on the radial coordinate. Its exterior
gravitational potential is determined by a function that has the dimension of
mass, whose value is proportional to mass contained in a unit of length of the
cylinder.

We now proceed by the same plan as we did in the quasi-spherical case. We can
freely move a circle of integration around each $(x, y)$ surface. We first
consider the hyperbolically symmetric case and we erect over a chosen circle a
solid column in the $z$ direction that contains a certain amount of rest mass.
Then we go over to the quasi-hyperbolic nonsymmetric space and erect a wiggly
column that contains the same amount of rest mass.

We will integrate over the interior of a circle in sheet 2 whose radius $u_0$
is, for the beginning, unknown. We only know that the radius must be smaller
than $S$, so that the integration region does not intersect the circle where
${\cal E} = 0$ (since, we recall, ${\cal E} = 0$ is the image of infinity, and
the integral over a region that includes ${\cal E} = 0$ would be infinite).
Thus, instead of (\ref{9.1}) and (\ref{11.1}) we have this time
 \begin{widetext}
\begin{eqnarray}\label{13.1}
\int_U {\rm d}_2 xy\ \frac 1 {{\cal E}^2} &=& \int_0^{2 \pi} {\rm d} \varphi
\int_0^{u_0} \frac {4 u S^2} {\left(u^2 - S^2\right)^2} {\rm d} u = 4 \pi\ \frac
{{u_0}^2} {S^2 - {u_0}^2}, \\
\int_U {\rm d}_2 xy\ \frac {{\cal E},_z} {{\cal E}^3} &=& \int_0^{2 \pi} {\rm d}
\varphi \int_0^{u_0} \frac {- 8 u S^2} {\left(u^2 - S^2\right)^3} \big(u \cos
\varphi P,_z + u \sin \varphi Q,_z + u^2 S,_z / (2S) + SS,_z/2\big)\ {\rm d} u =
\frac {4 \pi SS,_z {u_0}^2} {\left({u_0}^2 - S^2\right)^2}. \nonumber
\end{eqnarray}
 \end{widetext}
The first integral in (\ref{13.1}) will be independent of $S$ when $u_0$ is a
fixed multiple of $S$:
\begin{equation}\label{13.2}
u_0 = \beta S, \qquad \beta < 1.
\end{equation}
Then
\begin{eqnarray}\label{13.3}
\int_U {\rm d}_2 xy\ \frac 1 {{\cal E}^2} &=& 4 \pi\ \frac {\beta^2} {1 -
\beta^2}, \nonumber \\
\int_U {\rm d}_2 xy\ \frac {{\cal E},_z} {{\cal E}^3} &=& \frac {4 \pi (S,_z/S)
\beta^2} {\left(1 - \beta^2\right)^2}.
\end{eqnarray}
Instead of (\ref{11.2}) we now have:
\begin{eqnarray}\label{13.4}
{\cal M} &=& \frac {\beta^2} {1 - \beta^2} \int_{z_0}^z {\rm d} u \frac
{M,_u(u)} {\sqrt{2E - 1}} \nonumber \\
&-& \frac {\beta^2} {\left(1 - \beta^2\right)^2} \int_{z_0}^z {\rm d} u \frac
{3MS,_u} {S \sqrt{2E - 1}}.
\end{eqnarray}

The meaning of the limits of integration has to be explained here. In the
quasi-spherical case, and in the spherically symmetric LT subcase, one usually
assumes that each space of constant $t$ has its center of symmetry, where $M = 0
= R$. As explained at the beginning of Sec. \ref{sphermass}, this is an
additional assumption -- the center of symmetry need not belong to the
spacetime. But the center of symmetry, or origin, is the natural reference point
at which the mass function has zero value. In the quasi-hyperbolic case now
considered, a similar role is played by the set $M = 0$, so we will assume that
this set exists. Again, as mentioned earlier, this set is a 2-dimensional
surface in each space of constant $t$, and not a single point.

With this assumption made, $z_0$ in (\ref{13.4}) will be the value of $z$ at
which $M(z_0) = 0$. In addition, we assume that $(M,_u/\sqrt{2E - 1})$ and
$[S,_u/(S \sqrt{2E - 1})]$ are finite at $z = z_0$ and in a neighbourhood of
$z_0$, so that both integrals in (\ref{13.4}) tend to zero as $z \to z_0$.

These equations are very similar to the corresponding ones in the
quasi-spherical case, (\ref{11.1}) -- (\ref{11.2}), so one is tempted to
interpret $M$, by analogy with that case, as a quantity proportional to the
active gravitational mass contained within a solid (wiggly) tube with circular
sections, the radius of a circular section at $z = z_1$ being $\beta S(z_1)$.
The base of the tube is in the surface $(t = t_0 =$ const, $z)$, its coordinate
height is $(z - z_0)$, and its top is at $(t, z) = (t_0, z_0)$. There is no
problem with this interpretation in the hyperbolically symmetric case, where
$S,_u = 0$ and the second integral in (\ref{13.4}) vanishes.

However, there is a significant difficulty when going over to the
quasi-hyperbolic case. In the quasi-spherical case we were free to take the
limit of the integral extending over the whole sphere, which was $\beta \to
\infty$. Here, the integral over the whole hyperboloid would correspond to
$\beta \to 1$, and in this limit all the integrals (\ref{13.1}) -- (\ref{13.4})
become infinite. Worse still, the contribution from the dipole -- the second
integral in (\ref{13.4}) -- tends to infinity {\it faster} than the monopole
component (the first integral), while in the spherical case increasing the area
of integration caused decreasing the influence of the dipole.

We can do another operation on (\ref{13.4}) that will shed some light on the
meaning of $M$. The volume of the region containing the rest mass ${\cal M}$ is
\begin{equation}\label{13.5}
{\cal V} = \int_{z_0}^z {\rm d} u \int_U {\rm d}_2 xy \sqrt{\left|g_3(t, u, x,
y)\right|},
\end{equation}
where $g_3$ is the determinant of the metric of the 3-dimensional subspace $t =$
constant of (\ref{2.1}), thus
\begin{equation}\label{13.6}
{\cal V} = \int_{z_0}^z {\rm d} u \int_U {\rm d}_2 xy \frac {R^2 \left(R,_u - R
{\cal E},_u / {\cal E}\right)} {\sqrt{2E -1} {\cal E}^2}.
\end{equation}
This has the same structure as the integral representing ${\cal M}$. Since $R$
and $E$ do not depend on $x$ and $y$, the integration with respect to $(x, y)$
can be carried out, and by (\ref{13.1}) -- (\ref{13.3}) we get
\begin{eqnarray}\label{13.7}
{\cal V} &=& \frac {4 \pi \beta^2} {1 - \beta^2} \int_{z_0}^z \frac {R^2 R,_u}
{\sqrt{2E - 1}} {\rm d} u \nonumber \\
&-& \frac {4 \pi \beta^2} {\left(1 - \beta^2\right)^2} \int_{z_0}^z \frac {R^3
S,_u} {S \sqrt{2E - 1}} {\rm d} u.
\end{eqnarray}
For the ratio ${\cal M} / {\cal V}$ we now calculate two consecutive limits:
first $\beta \to 1$, to cover the whole of each $z =$ constant hyperboloid, and
then $z \to z_{\infty}$, where $z_{\infty}$ is the value of $z$ at which $R \to
\infty$, to cover the whole $t =$ constant space. After taking the first limit,
we get
\begin{equation}\label{13.8}
\lim_{\beta \to 1} \frac {\cal M} {\cal V} = \left.\left[\int_{z_0}^z \frac {3 M
S,_u} {S \sqrt{2E - 1}} {\rm d} u\right]\right/ \left[4 \pi \int_{z_0}^z \frac
{R^3 S,_u} {S \sqrt{2E - 1}} {\rm d} u\right].
\end{equation}
Since in general (apart from special forms of the functions involved) both the
numerator and the denominator above become infinite when $z \to z_{\infty}$, we
apply the de l'H\^{o}pital rule and obtain
\begin{equation}\label{13.9}
\lim_{z \to z_{\infty}} \lim_{\beta \to 1} \frac {\cal M} {\cal V} = \lim_{z \to
z_{\infty}} \frac {3 M} {4 \pi R^3}.
\end{equation}
The l.h.s. of the above is the global average of rest mass ${\cal M}$ per
volume. The r.h.s. looks very much like the same type of global average for the
active gravitational mass $M$, except that it is taken with respect to a flat
3-space.

A very similar result follows when we take $z \to z_0$ instead of $z \to
z_{\infty}$ in (\ref{13.8}). Then both the numerator and denominator tend to
zero and we obtain
\begin{equation}\label{13.10}
\lim_{z \to z_0} \lim_{\beta \to 1} \frac {\cal M} {\cal V} = \lim_{z \to z_0}
\frac {3 M} {4 \pi R^3}.
\end{equation}
On the l.h.s here we have a global average of ${\cal M} / {\cal V}$ over the
$(x, y)$ surface taken at the value of $z$ at which $M = 0$.

Equation (\ref{13.10}) results also when the integrals in (\ref{13.4}) --
(\ref{13.8}) are taken over the interval $[z_1, z_2]$, where $z_0 < z_1 < z_2 <
z_{\infty}$, and then the limit $z_2 \to z_1$ is taken.

All the calculations in this section were done in sheet 2 of the $(x, y)$ map.
The corresponding results for sheet 1 are obtained by taking all integrals with
respect to $u$ over the interval $[u_0, \infty)$ (with $u_0 > S$ now) instead of
$[0, u_0]$, substituting $1/u_0$ for $u_0$ in (\ref{13.1}), and $1/\beta$ for
$\beta$ (with the new $\beta$ obeying $\beta > 1$) in (\ref{13.3}), (\ref{13.4})
and (\ref{13.7}). Equations (\ref{13.8}) -- (\ref{13.10}) do not change.

The meaning of the limits on the r.h. sides of (\ref{13.9}) and (\ref{13.10})
requires further investigation. Note that they arise from the dipole
contributions to mass and volume.

\section{Summary}\label{summary}

\setcounter{equation}{0}

The aim of this paper was to clarify the geometrical structure of the
quasi-hyperbolic Szekeres models \cite{Szek1975a,Szek1975b} given by (\ref{2.7})
-- (\ref{2.9}), and of the associated hyperbolically symmetric dust model given
by (\ref{5.1}). The main results achieved are the following:

\begin{enumerate}
\item Although there exists no origin, where $R$ would be zero permanently, a
set where $M = 0$ can exist. At this location, $R,_t$ is constant (section
\ref{specprop}).

\item The whole spacetime is both future- and past- globally trapped (section
IV).

\item The geometrical interpretation of the $(x, y)$ coordinates in a
constant-$(t, z)$ surface was clarified in Sec. \ref{intcoord}. Contrary to an
earlier claim \cite{HeKr2008}, this surface consists of just one sheet, doubly
covered by the $(x, y)$ map.

\item The geometries of the following surfaces for the metric (\ref{5.1}) were
shown in illustrations, all of them in Sec. \ref{subspgeom}:

(a) $z = $ constant, $\varphi = 0$ for (\ref{5.1}) in Figs. \ref{constzembed} --
\ref{hypsurface}.

(b) The collection of $R(t, z)$ curves in Fig. \ref{evolutions}.

(c) $t = $ constant, $\varphi = 0$ for (\ref{5.1}) in Figs. \ref{multisheet} and
\ref{multisheetexp}.

It turned out that the surfaces listed under (a) are locally isometric to
ordinary surfaces of revolution in the Euclidean space (in special cases to a
plane and a cone) when $E \geq 1$, but cover the latter an infinite number of
times. When $1/2 < E < 1$, they cannot be embedded in a Euclidean space even
locally. The values $E \leq 1/2$ are prohibited by the spacetime signature.

The time evolution of the surfaces under 3(c) was illustrated in Figs.
\ref{evolwithedge} and \ref{evolwithnoedge}.

\item For the general metric (\ref{2.7}) -- (\ref{2.8}), the geometry of the
surfaces $t =$ constant, $\varphi = 0$ was investigated in Sec. \ref{diffshape}
and shown in Fig. \ref{concentric}. The other surfaces listed above are the same
as in the hyperbolically symmetric case (\ref{5.1}).

\item In Secs. \ref{sphmassinterpr} -- \ref{genqsphermass} a detailed analysis
was carried out of the relation between the mass function $M(z)$ and the sum of
rest masses in a volume ${\cal M}(z)$ in the quasi-spherical Szekeres model. The
function $M(z)$ represents the active gravitational mass within a sphere of
coordinate radius $z$, while ${\cal M}(z)$ is the sum of rest masses of
particles contained in the same volume:
\begin{equation}\label{14.1}
{\cal M} = \int_{\cal V} \sqrt{|g_3|} \rho {\rm d}_3 x,
\end{equation}
where ${\cal V}$ is any volume in a space of constant $t = t_0$, $g_3$ is the
determinant of the metric in that space and $\rho$ is the mass density at $t =
t_0$. The relation (\ref{8.3}) follows in the limit when ${\cal V}$ is the
volume of the whole space $t = t_0$. The calculations in Secs.
\ref{sphmassinterpr} -- \ref{genqsphermass} demonstrated how to calculate
(\ref{14.1}) within various relevant volumes.

\item In Secs. \ref{hypsym} and \ref{quasihyp} calculations analogous to those
from Secs. \ref{sphmassinterpr} -- \ref{genqsphermass} were carried out for the
quasi-hyperbolic Szekeres models. The aim was to interpret the function $M(z)$
in this case by identifying the volume in which the active gravitational mass is
contained. Integrals analogous to (\ref{14.1}) can be calculated, but the full
analogy with the quasi-spherical models follows only in the (hyperbolically)
symmetric case. In the general case, terms arising from the dipole component of
the mass distribution cause difficulties that were not fully resolved. It has
only been demonstrated that the average value of ${\cal M} / {\cal V}$ over the
whole space $t = t_0$ is determined by the average value of $M / V_0$, where
$V_0$ is the flat space limit of ${\cal V}$.

This problem requires further investigation, but it is hoped that the results
achieved here will be of use for that purpose.
\end{enumerate}

\appendix

\section{The curvature tensor for the metric (\ref{5.1})}\label{appsymm}

\setcounter{equation}{0}

The formulae given below are the tetrad components of the curvature tensor for
the metric (\ref{5.1}). The tetrad is the orthonormal one given by
\begin{eqnarray}\label{a.1}
{\rm e}^0 &=& {\rm d} t, \qquad {\rm e}^1 = \frac {R,_z} {\sqrt{2E - 1}}\ {\rm
d} z, \nonumber \\
{\rm e}^2 &=& R {\rm d} \vartheta, \quad {\rm e}^3 = R \sinh \vartheta {\rm d}
\varphi,
\end{eqnarray}
with the labeling of coordinates $(t, r, \vartheta, \varphi) = (x^0, x^1, x^2,
x^3)$. The components given below are scalars, so any scalar polynomial in
curvature components will be fully determined by them. Since they do not depend
on $\vartheta$, they have no singularity caused by any special value of
$\vartheta$. $\square$

\begin{eqnarray}
&& R_{0101} = \frac {2M} {R^3} - \frac {M,_z} {R^2 F}, \label{a.2} \\
&& R_{0202} = R_{0303} = \frac 1 2\ R_{2323} = - \frac M {R^3}, \label{a.3} \\
&& R_{1212} = R_{1313} = \frac M {R^3} - \frac {M,_z} {R^2 F}. \label{a.4}
\end{eqnarray}

The formulae in both appendices were calculated by the algebraic program
Ortocartan \cite{Kras2001, KrPe2000}.

\section{The curvature tensor for the metric (\ref{2.8})}\label{appnonsymm}

\setcounter{equation}{0}

The formulae given below are the tetrad components of the curvature tensor for
the metric (\ref{2.8}) with $\varepsilon = -1$. The tetrad is the orthonormal
one given by
\begin{eqnarray}\label{b.1}
{\rm e}^0 &=& {\rm d} t, \quad \quad {\rm e}^1 = \frac F {\sqrt{2E - 1}}\ {\rm
d} z, \nonumber \\
{\rm e}^2 &=& \frac R {\cal E}\ {\rm d} x, \quad {\rm e}^3 =  \frac R {\cal E}\
{\rm d} y,
\end{eqnarray}
with the labeling of coordinates $(t, z, x, y) = (x^0, x^1, x^2, x^3)$, where
${\cal E}$ is given by (\ref{2.7}) and
\begin{equation}\label{b.2}
F \df R,_z - R {\cal E},_z / {\cal E}.
\end{equation}
The components given below are scalars, so any scalar polynomial in curvature
components will be fully determined by them.
\begin{eqnarray}
&& R_{0101} = \frac {2M} {R^3} + \frac {3M {\cal E},_z} {R^2 {\cal E} F} - \frac
{M,_z} {R^2 F}, \label{b.3} \\
&& R_{0202} = R_{0303} = \frac 1 2\ R_{2323} = - \frac M {R^3}, \label{b.4} \\
&& R_{1212} = R_{1313} = \frac M {R^3} + \frac {3M {\cal E},_z} {R^2 {\cal E} F}
- \frac {M,_z} {R^2 F}. \label{b.5}
\end{eqnarray}
Note that these reproduce (\ref{a.2}) -- (\ref{a.4}) when ${\cal E},_z = 0$.

We wish to find out whether the sets ${\cal E} = 0$ and ${\cal E} \to \infty$
are curvature singularities. For this purpose it is useful to introduce the
coordinates $(\vartheta, \varphi)$ by (\ref{5.3}). Since the quantities
(\ref{b.3}) -- (\ref{b.5}) are scalars, we only need to substitute (\ref{5.3})
in them. The two suspected sets become $\vartheta \to \infty$ and $\vartheta =
0$, respectively. After the transformation we have
\begin{eqnarray}
&& {\cal E} = \frac S {2 \sinh^2(\vartheta/2)}, \label{b.6} \\
&& {\cal E},_z = \frac {S,_z} {2 \sinh^2(\vartheta/2)} \left[1 - 2 \cosh^2
(\vartheta/2)\right] \nonumber \\
&& \ \ \ \ \ - \coth (\vartheta / 2) \left(P,_z \cos \varphi + Q,_z \sin
\varphi\right). \label{b.7}
\end{eqnarray}

The only quantity in (\ref{b.3}) -- (\ref{b.5}) that depends on $\vartheta$ is
${\cal E},_z / ({\cal E} F)$. Using (\ref{b.6}) -- (\ref{b.7}) we easily find

\begin{eqnarray}
&& \lim_{\vartheta \to \infty} \frac {{\cal E},_z} {{\cal E} F} = \frac 1 R,
\label{b.8} \\
&&  \lim_{\vartheta \to 0} \frac {{\cal E},_z} {{\cal E} F} = \frac 1 {R,_z +
RS,_z / S}. \label{b.9}
\end{eqnarray}
The loci where these can become infinite do not depend on $\vartheta$. Hence,
$\vartheta \to \infty$ and $\vartheta = 0$ are not curvature singularities, and
neither are ${\cal E} = 0$ or ${\cal E} \to \infty$. $\square$

\bigskip

{\bf Acknowledgements} The research for this paper was inspired by a
collaboration with Charles Hellaby, initiated in 2006 at the Department of
Mathematics and Applied Mathematics in Cape Town. It was supported by the Polish
Ministry of Education and Science grant no N N202 104 838.


\bigskip

 \begin{thebibliography}{99}
\bibitem{HeKr2008} C. Hellaby and A. Krasi\'nski, {\it Phys. Rev.} {\bf D77},
023529 (2008).

\bibitem{Kras2008} A. Krasi\'nski, {\it Phys. Rev.} {\bf D78}, 064038 (2008);
    $+$ erratum: {\it Phys. Rev.} {\bf D85}, 069903(E) (2012). Fully corrected
version: arxiv:0805.0529v4.

\bibitem{Szek1975a}  P. Szekeres, {\it Commun. Math. Phys.} {\bf 41}, 55 (1975).

\bibitem{Szek1975b} P. Szekeres, {\it Phys. Rev. D} {\bf 12}, 2941 (1975).

\bibitem{BoTo1976} W. B. Bonnor, N. Tomimura, {\it Mon. Not. Roy. Astr.
Soc.} {\bf 175}, 85 (1976).
     %
\bibitem{GoWa1982a} S. W. Goode and J. Wainwright, {\it Mon. Not. Roy. Astr.
Soc.} {\bf 198}, 83 (1982).

\bibitem{GoWa1982b} S. W. Goode and J. Wainwright, {\it Phys. Rev.} {\bf D26},
3315 (1982).

\bibitem{Bonn1976a}  W.B. Bonnor, {\it Nature} {\bf 263}, 301 (1976).
     %
\bibitem{Bonn1976b}  W.B. Bonnor, {\it Commun. Math. Phys.} {\bf 51},
191-9 (1976).
     %
\bibitem{BoST1977} W. B. Bonnor, A. H. Sulaiman and N. Tomimura, {\it
Gen. Relativ. Gravit.} {\bf 8}, 549 (1977).

\bibitem{DeSo1985} M.M. de Souza, {\it Rev. Bras. Fiz.} {\bf 15}, 379 (1985).

\bibitem{Bonn1986} W. B. Bonnor, {\it Class. Quant. Grav.} {\bf 3}, 495 (1986).

\bibitem{BoPu1987} W. B. Bonnor, D. J. R. Pugh, {\it South Afr. J. Phys.}
{\bf 10}, 169 (1987).

\bibitem{Szek1980} P. Szekeres, in: {\it Gravitational radiation, collapsed
objects and exact solutions}. Edited by C. Edwards. Springer (Lecture Notes vol.
124), New York 1980, p. 477.
     %
\bibitem{Bole2006} K. Bolejko, {\it Phys. Rev.} {\bf D 73}, 123508 (2006).
     %
\bibitem{HeKr2002}  C. Hellaby and A. Krasinski, {\it Phys. Rev. D} {\bf 66},
    084011, (2002).
     %
\bibitem{PlKr2006} J. Pleba\'nski and A. Krasi\'nski, {\it An Introduction to
General Relativity and Cosmology} (Cambridge University Press, Cambridge,
England, 2006), ISBN 0-521-85623-X.
     %
\bibitem{Kras1997} A. Krasi\'nski, {\it Inhomogeneous Cosmological
Models}, Cambridge U P (1997), ISBN 0 521 48180 5.

\bibitem{BKHC2010} K. Bolejko, A. Krasi\'nski, C. Hellaby, and M.-N.
C\'el\'erier, {\it Structures in the Universe by exact methods -- formation,
evolution, interactions} (Cambridge University Press, Cambridge, England, 2006),
ISBN 978-0-521-76914-3.

\bibitem{KrBo2012} A. Krasi\'nski and K. Bolejko, {\it Phys. Rev. D} {\bf 85},
124016 (2012).

\bibitem{KaSa1966} R. Kantowski R. K. and Sachs, {\it J. Math. Phys.}
{\bf 7}, 443 (1966).

\bibitem{Hell1996}  C. Hellaby, {\it J. Math. Phys.} {\bf 37}, 2892 (1996).

\bibitem{Elli1971} G. F. R. Ellis,  in: {\it Proceedings of the International
School of Physics `Enrico Fermi', Course 47: General Relativity and Cosmology}.
Edited by R. K. Sachs. Academic Press, New York and London 1971, pp. 104 -- 182;
reprinted, with historical comments, in {\it Gen. Relativ. Gravit.} {\bf 41},
575 (2009).

\bibitem{RoNo1968} H. P. Robertson and T. W. Noonan (1968). {\it Relativity and
Cosmology}. W. B. Saunders Company, Philadelphia -- London -- Toronto, p. 374 --
378.

\bibitem{Kras1989} A. Krasi\'nski, {\it J. Math. Phys.} {\bf 30}, 433 (1989).

\bibitem{MuHe2001} N. Mustapha and C. Hellaby, {\it Gen. Relativ. Gravit.} {\bf 33},
455 (2001).

\bibitem{Bole2009} K. Bolejko, {\it Gen. Relativ. Gravit.} {\bf 41}, 1585 (2009).

\bibitem{Hell1987} C. Hellaby, {\it Class. Quant. Grav.} {\bf 4}, 635 (1987).

\bibitem{Kras2001} A. Krasi\'nski, {\it Gen. Relativ. Gravit.} {\bf 33}, 145 (2001).

\bibitem{KrPe2000} A. Krasi\'nski, M. Perkowski, {\it The system ORTOCARTAN --
user's manual}. Fifth edition, Warsaw 2000.
 \end{thebibliography}
 \end{document}